\documentclass[prd,twocolumn,superscriptaddress,floatfix,showpacs,nofootinbib]{revtex4-2}
\usepackage{amsfonts}
\usepackage{amsmath}
\usepackage{amssymb}
\usepackage{xcolor}
\usepackage{graphicx}
\usepackage[normalem]{ulem}
\font\mybb=msbm10 at 10pt
\def\bb#1{\hbox{\mybb#1}}
\def\be{\begin{equation}}
\def\ee{\end{equation}}
\def\bea{\begin{eqnarray}}
\def\eea{\end{eqnarray}}
\newcommand{\p}[1]{(\ref{#1})}
\begin{document}
\begin{flushright}
%V1:26/05/2026 $\quad$ V2:03/06/2026 $\quad$ V3:19/07/2026 $\quad$ 
%arXiv:2605.27754[hep-th]
Phys. Rev {\bf D114} (2026) n4, 046023
\end{flushright}
\bigskip
\title{Twistor approach to classical and quantum D$0$--brane}
\author{Igor Bandos}
\email{igor.bandos@ehu.eus}
\affiliation{Department of Physics and EHU Quantum Center, University of the Basque Country UPV/EHU,
P.O. Box 644, 48080 Bilbao, Spain,}
\affiliation{IKERBASQUE, Basque Foundation for Science,
48011, Bilbao, Spain, }

\author{Mirian Tsulaia}
\email{mirian.tsulaia@oist.jp}
\affiliation{Okinawa Institute of Science and Technology,
1919-1 Tancha, Onna-son, Okinawa 904-0495, Japan}

\bigskip

\begin{abstract}
We develop the (super)twistor approach to D$0$--brane, which is the massive type IIA superparticle in ten dimensional spacetime. The basic variables
are {{hexadecuplet}} of constrained $OSp(32|1)$  supertwistors similar but not identical to the ones which have been used for the description of 11D massless superparticle, also known as M$0$--brane. We show how the  constrained supertwistor formulation is related to  the spinor moving frame approach to D$0$--brane. We perform the quantization of the model by two different methods and discuss the relation of the results with the quantization  of the spinor moving frame formulation of D$0$--brane. The quantum state spectrum describes the massive counterpart of the linearized type IIA supergravity.

\end{abstract}

\maketitle
\begin{widetext}
\tableofcontents

\section{Introduction}

It is well known that
dynamical nonperturbative objects, $p$-branes and particularly Dirichlet p--branes or
D$p$-- branes,  being  an integral  part of String Theory, are essential  for its overall consistency and for establishing various strong/weak coupling dualities between  different branches of String/M-theory.
 D$0$-branes or  Dirichlet particles are the simplest and yet nontrivial examples of D$p$--branes
 belonging to a nonperturbative spectrum of the type
IIA string theory. Its standard (Brink--Schwarz type) action {{can be obtained}} by dimensional reduction of the standard action of M$0$--brane (or M-wave), which is the massless $D=11$ superparticle \cite{Bergshoeff:1996tu}. This fact is one of the manifestations of
 the duality between type IIA superstring
string theory and other ''corners'' of M--theory, which can be considered as
de-compactification limit of the former  in its strong coupling regime \cite{Witten:1995im}.
The quantum state spectrum of the D$0$-brane  was shown to be the linearized
 massive type IIA supermultiplet,
with $128$ bosonic and $128$ fermionic degrees of freedom,
which can be obtained
from the eleven dimensional supergravity as a nontrivial mode in the Kaluza-Klein dimensional reduction  \cite{Bandos:2025pxv}.

The above mentioned quantum state spectrum of  D$0$--brane was found and studied in
\cite{Bandos:2025pxv} devoted to the quantization of  this dynamical system in the so--called spinor moving frame formalism   (see \cite{Kallosh:1997nr} for an earlier discussion of quantization of a D$0$--brane). This formalism  was proposed in \cite{Bandos:2000tg} and used there as the basis for generalized action principle and superembedding approach to this object. Notice that the action for multiple D$0$--branes
(mD$0$ system) recently constructed and elaborated in
\cite{Bandos:2022uoz, Bandos:2022dpx} is presently known only in the frame of the spinor moving frame approach.

In this paper we  develop a twistor approach to D$0$--brane in the classical and quantum domain.
As it could have been  expected
from studies
of similar systems in $D=4$
(see for example \cite{Ferber:1977qx, Shirafuji:1983zd,Eisenberg:1988nt, Bandos:1999qf, Bandos:2005mb}),
the formulation in terms of twistorial variables considerably simplifies
both the Hamiltonian analysis and the quantization procedure.
Furthermore, the results of such quantization provided us with a one-particle counterpart of the on-shell superamplitudes \cite{Brandhuber:2008pf,Arkani-Hamed:2008gz}   being an important tool for reaching an impressive progress in multiloop calculations in maximal $D=4, {\cal N}=4$ supersymmetric gauge theory and  $D=4, {\cal N}=8$ supergravity \cite{Bern:2011qn,Elvang:2015rqa}.

The generalization of twistor approach \cite{Penrose:1967wn,Penrose:1972ia,Penrose:1986ca,Ferber:1977qx,Shirafuji:1983zd} to the case of  massive $D=4$ superparticle was elaborated in  \cite{Fedoruk:2003td,Bette:2004ip,Fedoruk:2005ks} under the name  of two--twistor formalism  (see e.g. \cite{deAzcarraga:2005ky,Fedoruk:2005bs,deAzcarraga:2008ik,deAzcarraga:2014hda,Uvarov:2019vmd,Kim:2021rda,Kim:2026opo,Kim:2026yqo} for further related studies). The amplitudes with $D=4$ massive particles in the frame of  $D=4$ spinor helicity formalism, which can be related to two--twistor approach,   was the subject of  \cite{Dittmaier:1998nn,Conde:2016izb,Arkani-Hamed:2017jhn}. The first stages towards generalization of these results for the case of amplitudes and superamplitudes of the processes involving D$0$--branes were discussed in \cite{Bandos:2025pxv} and \cite{Bandos:2026wyk}. We believe that the development of pure twistor description of D$0$--brane which we present below will be useful for further development of this line.

The paper is organized as follows.

In section \ref{D4}, as a warm-up,
we briefly review a
well known twistor--like model of massless
$D=4$ ${\cal N}=1$ superparticle \cite{Shirafuji:1983zd}, making emphasis on its re-formulation in terms of Ferber supertwistors \cite{Ferber:1977qx} and quantization in terms of supertwistor variables. We also discuss there the relation of Ferber--Schirafuji formulation with spinor moving frame or Lorentz harmonic approach to $D=4$ superparticle \cite{Bandos:1990ji} because this allows for the ten dimensional generalization which we  will use in section \ref{D0twistor}.

In section \ref{D0twistor} we develop the constrained supertwistor {{approach}} to D$0$--brane in the classical domain. Our starting point is the spinor moving frame approach for
 D$0$--brane
\cite{Bandos:2000tg, Bandos:2025pxv} which we review
in sec. \ref{D0-spinorMF}.
In sec. \ref{sec:lambda=vS} we introduce the constrained helicity spinors, related to the spinor moving frame variables by $SO(16)$ transformations. These are used as a basic element of the constrained $OSp(32|1)$ supertwistor approach. The action of D$0$--brane in this approach is obtained in sec.
\ref{sec:YM-SO16} where the gauge symmetries of this action and its relation to the so-called tensorial superspace are also discussed. Hamiltonian mechanics of the constrained supertwistor formulation of D$0$--brane is developed in sec. \ref{sec:0-Tw-Ham}.

In section \ref{quantization} we perform
the quantization of D$0$--brane in its constrained supertwistor formulation. First, in sec. \ref{QunatumD0-tw},  we do this by the method which have been used in \cite{Bandos:2006nr} for quantization of supertwistor formulation of massless $D=11$ superparticle (see \cite{Green:1999by} for similar method in the frame of light-cone quantization). The result clearly indicates that the quantum state spectrum of the D$0$--brane is given by
massive counterpart of  $D=10$ type IIA supergravity supermultiplet,  in agreement with the results of \cite{Bandos:2025pxv}. In sec. \ref{Quant=2} we perform an alternative quantization resulting in a multi--superfield description of quantum state vector of D$0$--brane. This allows us in sec. \ref{Quant=3} to reproduce the analytic on-shell superfield description of the quantum state vector of D$0$--brane obtained in \cite{Bandos:2025pxv}.

The conclusion and discussion on possible applications of our results can be found in
 sec. \ref{Conclusion}.

Appendices collect the description of our notation and some useful technicalities.

\section{Massless $D=4$ ${\cal N}=1$ Superparticle. A Reminder}
\label{D4}
\setcounter{equation}0

Let us briefly recall the main features
of classical and quantum description
of massless $D=4$ ${\cal N} =1$ superparticle model
in terms of twistorial variables {\cite{Ferber:1977qx,Shirafuji:1983zd}.
The suitable classical action to start with is given
by \cite{Shirafuji:1983zd}

\be \label{d4action-1}
S =
\int\limits_{\mathcal{W}^1} \,   \Pi^{\alpha \dot \alpha}\lambda_\alpha \,
\bar \lambda_{\dot \alpha}=\int \, d \tau \,  \Pi_\tau^{\alpha \dot \alpha}\lambda_\alpha \,
\bar \lambda_{\dot \alpha},
\qquad  \qquad \alpha, \dot \alpha = 1,2,
\ee
where $\Pi^{\alpha \dot \alpha}$ is the pull--back of the Volkov -- Akulov one form
\be
\Pi^{\alpha \dot \alpha} = d x^{\alpha \dot \alpha} + id \theta^\alpha\,  \, \bar \theta^{\dot \alpha}
- i \theta^\alpha\,  \, d \bar \theta^{\dot \alpha}=  d \tau \,  \Pi_\tau^{\alpha \dot \alpha}
\ee
to the worldline ${\mathcal{W}^1}$ in the  superspace enlarged by bosonic spinor variables $\lambda^\alpha $ and $\lambda^\alpha$. This line is defined parametrically in terms of coordinate functions of the proper time
$\tau$,
\be\label{cZ=}
{\cal Z}^{\cal M}(\tau)= (x^{\alpha\dot \alpha} (\tau), \theta^\alpha(\tau), \bar \theta^{\dot \alpha}(\tau), \lambda^\alpha(\tau), \bar \lambda^{\dot \alpha}(\tau) ) \; . \qquad
\ee

Constructing the Hamiltonian mechanics starting from the action  \p{d4action-1},
one obtains the primary constraints
\be \label{constr-1}
\Phi_{\alpha \dot \alpha } = p_{\alpha \dot \alpha }
- \lambda_\alpha \bar \lambda_{\dot \alpha} =0,
\ee
\be \label{constr-2}
p_\alpha =0, \qquad p_{\dot \alpha} =0, \qquad
D_\alpha = - \pi_{\alpha} + i  p_{\alpha \dot \alpha }\bar \theta^{\dot \alpha}, \qquad
D_{\dot \alpha} = - \pi_{\dot \alpha} + i  p_{\alpha \dot \alpha } \theta^{ \alpha}
\ee
where $p_{\alpha \dot \alpha }$,
$p_\alpha$, $\bar p_{\dot \alpha}$,
$\pi_\alpha$ and $\bar \pi_{\dot \alpha}$ are momenta, canonically conjugated to $x_{\alpha \dot \alpha }$,
$\lambda^\alpha$, $\bar \lambda^{\dot \alpha}$,
$\theta^\alpha$ and $\bar \theta^{\dot \alpha}$.
After introducing the Poisson brackets
 between canonical variables
\be
[ p_{\alpha \dot \alpha }, x^{\beta \dot \beta } ]_{P.B.}=
\delta ^\beta_\alpha \delta^{\dot \beta}_{\dot \alpha}, \quad
[ p^\alpha, \lambda_\beta ]_{P.B.}  =
 \delta^\alpha_\beta, \quad
[ \bar p_{\dot \alpha}, \bar \lambda^{\dot \beta} ]_{P.B.}= \delta^{\dot \beta}_{\dot \alpha}, \quad
\{ \pi_\alpha,\theta^\beta \}_{P.B.}  =\delta_\alpha^\beta,
 \quad
\{ \bar \pi_{\dot \alpha},\bar \theta^{\dot \beta} \}_{P.B.}  =\delta_{\dot \alpha}^{\dot \beta}
\ee
one can see that the part of the
total set of ($8$ bosonic and $4$ fermionic)
constraints are of the second class
(see e.g. \cite{Bandos:1999qf} for details), which makes the quantization
procedure complicated.
In particular, $2$ bosonic and $2$
fermionic constraints are of the first class, the rest of the constraints are of the second class.
This makes the total number of the
independent bosonic phase space
variables\footnote{ The counting goes as follows.
 The  phase space contains $2\times (4+4)=16$ bosonic and
$2\times 4=8$ fermionic degrees of freedom.
Two bosonic  first class constraints eliminate  $4$, and six bosonic
second class constraints eliminate $6$.
degrees of freedom.
Similarly,
two fermionic first  class constraints eliminate $4$ and two fermionic second  class constraints eliminate $2$ degrees of freedom, thus leaving us with
six bosonic and two fermionic degrees of freedom.}
equal to $6$ and the number
of independent fermionic phase space variables equal to $2$.

Despite  its simplicity, the quantization of the model is nontrivial due to the presence of the second-class constraints.
A possible approach
that simplifies the problem
is
to
  change the variables to super-twistors
\be {\cal Z}_A= (\lambda_\alpha, {\bar \mu}^{\dot \alpha}, \chi)\; , \qquad
{\bar {\cal Z}}^A= (\mu^\alpha, {\bar \lambda}_{\dot \alpha}, \bar \chi)\; , \ee
the components of which are related to the original coordinate functions \eqref{cZ=} by the supersymmetric generalization  of the Penrose incidence relations \cite{Ferber:1977qx}
\be \label{4dtw}
\mu^\alpha = x^{\alpha \dot \alpha} \bar \lambda_{\dot \alpha} + i \, \theta^\alpha (\bar \theta^{\dot \beta} \bar \lambda_{\dot \beta}),
\qquad
\bar \mu^{\dot \alpha}=
x^{\alpha \dot \alpha}  \lambda_{ \alpha} + i \, \bar \theta^{\dot \alpha} ( \theta^{\dot \beta}  \lambda_{ \beta}),
\qquad
 \chi = \theta \lambda, \qquad
 \bar \chi = \bar \theta^{\dot \alpha} \bar \lambda_{\dot \alpha}\;.
\ee
The action \p{d4action-1}
can now be written in terms of the supertwistor variables,
\be \label{4daction-tw}
S = -\int\limits_{\mathcal{W}^1} \bar {\cal Z}^A d {\cal Z}_A \; ,
\ee
without any reference to $x^{\alpha \dot \alpha}$ $ \theta^\alpha$ and  $\bar \theta^{\dot \alpha} $,
provided these variables obey   the following  constraint
\be \label{4dtw-c}
{\cal Z}^A {\cal Z}_A = \mu^\alpha \lambda_\alpha - \bar \mu^{\dot \alpha} \bar \lambda_{\dot \alpha} + 2i \bar \chi \chi =0\; .
\ee
This is correct because   the general solution of Eq. \eqref{4dtw-c} is
 provided by Eqs. \p{4dtw}.

In the Hamiltonian formalism the constraint \eqref{4dtw-c} reflects the presence (generates) the $U(1)$ gauge symmetry of the system, which  makes the number of independent phase space variables
 equal to $6$ in agreement with our previous counting.

 To perform  quantization
of the system defined by the free twistorial action
\p{4daction-tw} restricted by \p{4dtw-c} is easy. Indeed,  after reducing the phase space by accounting for explicitly solved bosonic second class constraints and passing to the standard Dirac brackets for the fermionic second class constraints, one finds that the only condition
that the state vector has to satisfy, is the quantum version of the constraint
\p{4dtw-c}. Replacing the canonical Dirac  brackets
\be
[{ \mu}_{ \alpha}, { \lambda}^{ \beta}]_{D.B.}= \delta_{ \alpha}^{ \beta},
\qquad
[{\bar \mu}_{\dot \alpha}, {\bar \lambda}^{\dot \beta}]_{D.B.}= \delta_{\dot \alpha}^{\dot \beta}, \qquad
\{ \chi, \bar \chi \}_{D.B.} = -\frac{i}{2}
\ee
with commutators and anticommutator according to (updated) Dirac rules,
$ [ \ldots , \ldots  \}_{D.B.} \rightarrow \,\frac 1 i \, [  \ldots ,  \ldots  \}$,
and passing to the coordinate ($\lambda$-) representation for bosonic and holomorphic
($\chi$-) representation for fermionic variables,  one finds that the quantum version of \eqref{4dtw-c} reads
 \be \label{4dqc}
\left ( \lambda_\alpha \frac{\partial}{\partial \lambda_\alpha} -
\bar \lambda_{\dot \alpha} \frac{\partial}{\partial {\bar \lambda}_{\dot \alpha}} + \chi \frac{\partial}{\partial \chi} -s\right ) \Phi (\lambda, \bar \lambda, \chi) =0\; .
\ee
The constant $s$ in this equation
is due to the ordering ambiguity. It is quantized and has the meaning of (super)helicity of the
massless supermultiplet describing the quantum state spectrum of the superparticle
 (see e.g. \cite{Bandos:1999qf}).

In the simplest case of $s=0$, the solution of Eq. \eqref{4dqc} is a chiral on-shell superfield
\be \label{4dsuperfield}
\Phi(\lambda, \bar{\lambda}, \chi ) = \phi(\lambda) + i \chi \, \bar \psi (\lambda), \qquad
\ee
the components of which describe the general solutions $\phi (p)\vert_{p^2=0}$, $\bar \psi_{\dot \alpha} (p) = \bar{\lambda}_{\dot \alpha}\, \bar \psi (\lambda)$  of the linearized equations of motion for scalar supermultiplet fields, $p^2\, \phi (p)=0$ and $p_{\alpha \dot \alpha}\bar{\psi}^{\dot \alpha} =0\;$. This solution is based on solving  $p^2=0$  by the Cartan--Penrose representation   $p_{\alpha \dot \alpha}
= \lambda_\alpha \bar \lambda_{\dot \alpha}$.

The spinor moving frame description of $D=4$ superparticle \cite{Bandos:1990ji} is closely related to the above Ferber-Schirafuji supertwistor formalism described above, but allows for an easy generalization for higher dimensions and for supersymmetric extended objects, including  $D=4$ and $D=10$ superstrings \cite{Bandos:1992np,Bandos:1992ze}, $D=11$ supermembrane \cite{Bandos:1992hu,Bandos:1993yc}.and other p--branes \cite{Bandos:1994eu,Bandos:2006wb}.  We briefly describe it here as its generalization for the case of
$D=10$ D$0$--brane \cite{Bandos:2000tg,Bandos:2025pxv}, which we review in sec. \ref{D0-spinorMF}, will be used as a starting point to develop supertwistor approach to this object beginning sec \ref{sec:lambda=vS}.
%%%%%%%%%%%%%

%%%%%%%%
To make its appearance natural, let us consider a D=4 massless particle in a special Lorentz  frame in which its light-like momentum reads
$
P_{(a)} = (1,0, 0,1) \rho^{++}(\tau)
$,
where $\rho^{++}(\tau)$ is the energy of the particle.
In an arbitrary frame the momentum is
$
p_\mu = U_\mu^{(a)} \, P_{(a)} \equiv u_\mu^{--} \rho^{++}
$,
where
\be \label{4dmfm}
U_\mu^a = \left (  \frac{1}{2}(u_\mu^{++} + u_\mu^{--}), \,
u_\mu^i, \,  \frac{1}{2}(u_\mu^{++} - u_\mu^{--})  \right) \in {{SO(1,3)}}, \qquad i=1,2
\ee
$U_\mu^a(\tau)$ is a {{Lorentz}} group valued matrix constructed from
two light--like vectors $u_\mu^{++}$, $u_\mu^{--}$ and {{spacelike}} vectors $u_\mu^i$  orthogonal to these and among themselves.
This follows
from  \eqref{4dmfm}, which is tantamount to saying that
$
U_\mu^a \, U^{\mu b} = \eta^{ab}
$ i.e. {{
\bea u_\mu^{++}u^{\mu++}=0\; , \qquad u_\mu^{--}u^{\mu--}=0\; , \qquad u_\mu^{--}u^{\mu++}=2\; , \nonumber \\  \qquad u_\mu^{i}u^{\mu\pm\pm}=0\; , \qquad  u_\mu^{i}u^{\mu j}=-\delta^{ij}\; . \qquad \eea }}
The above equations, but with $i=1,..., D-2$, are actually valid for any D. In $D=4$ it is convenient to represent $D-2=2$ orthogonal moving frame vectors
by complex conjugate $u_\mu^{+-}= \frac{1}{2} (u_\mu^1 + i u_\mu^2)$ and
$u_\mu^{-+}= \frac{1}{2} (u_\mu^1 - i u_\mu^2)$.

The $D=4$  spinor moving frame variables  consist of a pair of complex bosonic spinors $v_\alpha^\pm$ normalized by
\be
v^{\alpha -} v_\alpha^+ =1, \qquad \bar v^{\dot \alpha -} \bar v_{\dot \alpha}^+ =1
\ee
which are "square root" from the  above  moving frame variables
in the sense of \be\label{u--=ll=4D}
u_\mu^{--} = v^- \sigma_\mu \bar v^{-}, \qquad
u_\mu^{++} = v^+ \sigma_\mu \bar v^{+}, \qquad
u_\mu^{+-} = v^+ \sigma_\mu \bar v^{-}, \qquad
u_\mu^{-+} = v^- \sigma_\mu \bar v^{+}\; . \qquad
\ee
This is tantamount to saying that   spinor moving frame matrix
 $
v_\alpha^{(\beta)}= (v_\alpha^+,v_\alpha^-) \in SL(2,C)
$
 is a ''square root'' of the moving frame matrix \p{4dmfm}.

Using \eqref{u--=ll=4D} to write  the action \p{d4action-1}
in terms of the spinor moving frame variables,
\be \label{d4action-2}
S = \int\limits_{\mathcal{W}^1} \rho^{++} u_\mu^{--} \Pi^\mu =
\int d \tau  \rho^{++} (v^-\sigma_\mu \bar v^-) \Pi_\tau^\mu
\ee
we see that it is related to the Ferber--{{Schirafuji}} action $\int d \tau (\lambda \sigma_\mu \bar \lambda) \Pi_\tau^\mu$ by simple redefinition of the bosonic spinor variables  $\lambda_\alpha = \sqrt{\rho^{++}} v_\alpha^{-}$ and
$\bar \lambda_{\dot \alpha} = \sqrt{\rho^{++}} \bar v^+_{\dot \alpha}$.

To resume, the action the massless $D=4$ ${\cal N}=1$ superparticle can be reformulated in  terms of (super)twistor variables which simplifies essentially the quantization. In the next section III we will develop, on the basis of spinor moving frame formulation of D$0$--brane, its formulation in terms of constrained supertwistors. The relations of these latter with spinor moving frame variable
is not so straightforward as in four dimensional case and requires the use of
$SO(16)$, $SU(8)$ and also of
a more exotic symmetry realized by Stueckelberg mechanism.

\section{
Supertwistor approach to D$0$-brane. Lagrangian and Hamiltonian mechanics
}
\label{D0twistor}

\subsection{Spinor moving frame formulation of D$0$--brane}

\label{D0-spinorMF}

We begin by reviewing the  generalization of the above derivation of spinor moving frame formulation for a more complicated case of massive  superparticle in ten dimensions \cite{Bandos:2000tg} which is to say for Dirichlet superparticle or D$0$--brane \cite{Bergshoeff:1996tu}.
This object lives in
 $D=10$ type IIA superspace with coordinates
\be\label{ZM} Z^M=( x^\mu, \theta^{\alpha 1}, \theta_{\alpha}^2) \; \qquad \mu=0,\ldots,9\; , \qquad \alpha =1,\ldots, 16 \;  \qquad  \ee
and is invariant (up to boundary terms) under the rigid supersymmetry transformations
\begin{equation}\label{susy=IIA}
\delta x^\mu =i\theta^1\sigma^\mu \epsilon^1 +i\theta^2\tilde{\sigma}^\mu \epsilon^2 \; , \qquad \delta  \theta^{\alpha 1}= \epsilon^{\alpha 1}  \; , \qquad \delta  \theta^2_{\alpha}= \epsilon^2_{\alpha} \; . \qquad
\end{equation}
{{Here $\theta^1\sigma^\mu \epsilon^1=\theta^{1\alpha}\sigma^\mu_{\alpha\beta} \epsilon^{1\beta}$ and
$\theta^2\tilde{\sigma}^\mu \epsilon^2=\theta^2_{\alpha}\tilde{\sigma}^{\mu\, \alpha\beta}\epsilon^{2\beta}$,  where
$\sigma^\mu_{\alpha\beta}=\sigma^\mu_{\beta\alpha}$ and  $\tilde{\sigma}{}^{\mu\alpha\beta}=\tilde{\sigma}{}^{\mu\beta\alpha}$ are 10D generalizations of 4D relativistic Pauli matrices; they obey
\be
\sigma^\mu\tilde{\sigma}{}^\nu+ \sigma^\nu\tilde{\sigma}{}^\mu = \eta^{\mu\nu} {\bb I}_{16\times 16}\; , \qquad  \eta_{\mu \nu}= diag(1,-1,...,-1).
\ee
}}

The first order form of the standard D$0$--brane action
\cite{Bergshoeff:1996tu} reads
\be\label{SD0-standard} \int d\tau \left(p_\mu\Pi_\tau^\mu - \frac 1 2 e (p^2-m^2)\right) -im \int\limits_{\mathcal{W}^1} \left(\text{d} \theta^{\alpha 1}\theta_{\alpha}^{2} - \theta^{\alpha 1}\text{d}\theta_{\alpha}^{2} \right).  \ee
where $\Pi^\mu_\tau= \Pi^\mu/\text{d}\tau$ and $\Pi^\mu $ is type IIA counterpart of
the Volkov-Akulov $1$-form,
\begin{equation}
\Pi^\mu = \text{d}x^\mu -i\text{d}\theta^1\sigma^\mu \theta^1 -i\text{d}\theta^2\tilde{\sigma}^\mu \theta^2 =d\tau \Pi^\mu_\tau \; .
\end{equation}

The moving frame  action for  $D0$-- brane   \cite{Bandos:2000tg} can be written as
\begin{equation}\label{eq:LD0}
S^{\text{D}0} = \int\limits_{\mathcal{W}^1}{\cal L}_{1}^{\text{D}0}\equiv \int \text{d}\tau L^{\text{D}0} = m\int\limits_{\mathcal{W}^1}\Pi^\mu u^0_\mu -im \int\limits_{\mathcal{W}^1} \left(\text{d} \theta^{\alpha 1}\theta_{\alpha}^{2} - \theta^{\alpha 1}\text{d}\theta_{\alpha}^{2} \right)~, \qquad \qquad u^0_\mu u^{0\mu}=1\; ,
\end{equation}
where the unit timelike vector  $u^0_\mu$ is considered
as a part of the moving frame matrix
\bea\label{harmU=m}
(u_\mu^{0}, u_\mu^{I})\in \text{SO}(1,9) \;  \qquad I=1,\ldots,9\; . \qquad
 \eea
Eq. \eqref{harmU=m} is tantamount to stating that the moving frame vectors obey
\be
\label{u0u0=1}
u^{\mu 0}u_\mu^{0}=1\; , \qquad u^{\mu 0} u_\mu^{I}=0\; , \qquad u^{\mu I} u_\mu^{J}=-\delta ^{IJ},\;
\qquad
\epsilon^{\mu_1\ldots \mu_{10}}u_{\mu_1}^{0} u_{\mu_2}^{i_1}\ldots  u_{\mu_{10}}^{i_9} =
\epsilon^{i_1\ldots i_{9}}\;.
\ee

The action \eqref{eq:LD0}
can be obtained from  the standard D$0$--brane action \eqref{SD0-standard}
when  solving the mass-shell condition for auxiliary momentum variable, $p^2=m^2$,
by $p_\mu = m u^0_\mu$.

The spinor moving frame action is given by \eqref{eq:LD0} in which the moving frame vectors are considered to be composites of
the spinor moving frame  matrix
\be\label{v=inSpin}
v_\alpha{}^q \in \text{Spin}(1,9) \; ,  \qquad \alpha =1,\ldots, 16 \; ,  \qquad q =1,\ldots, 16
\ee
in the sense of the following constraints (see \cite{Bandos:2000tg,Bandos:2022uoz,Bandos:2022dpx} for more details)
\begin{eqnarray}\label{u0s=vv}
 u_\mu^{0} \sigma^\mu_{\alpha\beta}=v_\alpha{}^q v_\beta{}^q \; ,  \qquad \nonumber \\
u_\mu^{I} \sigma^\mu_{\alpha\beta}=v_\alpha{}^q \gamma^I_{qp}v_\beta{}^p \; , \qquad
 \\ \label{vtsv=}
v_{\alpha}^q \tilde{\sigma}{}_{\mu}^{\alpha\beta}v_{\beta}^p= u_\mu^{0} \delta_{qp}+u_\mu^{I} \gamma^I_{qp}\; ,  \qquad
\end{eqnarray}
where $\gamma^I_{qp}$ are 9d gamma matrices,
$I=1,\ldots,9$ {{obeying \begin{eqnarray}\label{gIgJ=dIJ=A}
\gamma^I_{qp} =\gamma^I_{pq} \; ,  \qquad  \gamma^I_{qq} =0 \; ,  \qquad  \gamma^I\gamma^J+\gamma^J\gamma^I=2\delta^{IJ} {\mathbb I}_{16\times 16} \; .  \qquad
\end{eqnarray}}} Taking the sigma--trace of  the first equation in
\eqref{u0s=vv}, or taking the trace of \eqref{vtsv=} (and taking into account that $\gamma^I_{qq}\equiv 0$), we find
\begin{equation}
u^{0 \mu} =\frac{1}{16}v_\alpha{}^q  \tilde{\sigma}{}^{\mu {\alpha\beta}} v_\beta{}^q  \; .
%=\frac{1}{16}v_q {}^\alpha \sigma^\mu_{\alpha\beta} v_q{}^{\beta}.
\end{equation}
Substituting this into the action
\p{eq:LD0},
one can see that it provides
a $D=10$ massive generalization
of the Ferber--{{Schirafuji}} action \eqref{d4action-1} for massless $D=4$ superparticle model considered in  section \ref{D4}.

The integration of the Lagrangian form ${\cal L}_1^{D0}$ in the action \eqref{eq:LD0} is over
the worldline in an enlarged superspace (Lorentz harmonic superspace) which can be defined parametrically
\be\label{WinLH=SSP}
{\cal W}^1 \in \Sigma^{(10+45|16+16)}: \qquad {\cal Z}^{\cal M}={\cal Z}^{\cal M}(\tau) = (x^\mu (\tau), \theta^{\alpha 1 } (\tau),  \theta_{\alpha}^2 (\tau),  v_\alpha{}^q(\tau))\;
\ee
with the use of coordinate functions ${\cal Z}^{\cal M}(\tau)$ of the proper time $\tau$. Let us stress that the set of these includes the spinor moving frame matrix \eqref{v=inSpin}.

The derivatives of the spinor moving frame matrix are expressed  in terms of SO(1,9) Cartan forms
\be\label{OmI=}
\Omega^{I}= u_\mu^{(0)} d u^{\mu I}, \qquad \Omega^{IJ}=u_\mu^{I} d u^{\mu J}
\ee
by
\cite{Bandos:2022uoz, Bandos:2022dpx}
\be\label{vdv=}
v_p{}^\alpha \text{d} v_\alpha{}^q
= - \frac{1}{4} \Omega^{IJ} \gamma^{IJ}_{pq} + \frac{1}{2}
 \Omega^{I} \gamma^{I}_{pq} .
\qquad
\ee
We have written this equation using $v_p{}^\alpha$ matrix which is the inverse to $v_\alpha{}^p$,
\be
v_p{}^\alpha v_\alpha{}^q=\delta^{pq}\; , \qquad  v_\alpha{}^qv_q{}^\beta=\delta_\alpha{}^\beta\;  \qquad
\ee
as this will be useful below. Notice that, as the charge conjugation matrix for $SO(9)$ group is symmetric and can be chosen to coincide with $\delta_{qp}$, the position of $Spin(9)$ indices is not important and we chose it from aesthetic and transparency reasons.

Of the Cartan forms \eqref{OmI=}, $\Omega^{IJ}$ transforms under the gauge $SO(9)$ symmetry of the action \eqref{eq:LD0} as a connection while $\Omega^{I}$ transforms covariantly under this symmetry.
These reflect
the fact that of $45$ degrees of freedom in moving frame matrix obeying \eqref{v=inSpin} (this is to say they  are restricted by constraints \eqref{u0s=vv} and \eqref{vtsv=}),
$36$ are pure gauge while $9$ are physical, the same as present in a timelike momentum of the massive $D=10$ particle.

The {\it covariant harmonic derivatives}, this is to say the vector fields (differential operators) which are dual to the Cartan forms \eqref{OmI=}, are
\bea\label{DIJ=}
D^{IJ}=\frac 12 v_\alpha^p\gamma^{IJ}_{pq}\frac {\partial } {\partial v_\alpha^p}\; , \qquad
%\\ \label{DI=}
D^{I}=\frac 12 v_\alpha^p\gamma^{I}_{pq}\frac {\partial } {\partial v_\alpha^p}\; . \qquad
\eea
These give zero when acting on the constraints \eqref{u0s=vv} and \eqref{vtsv=}.

In checking that \eqref{u0s=vv} and \eqref{vtsv=}  are in kernel of the operators \eqref{DIJ=}, one should consider the moving frame vectors to be composed from spinor moving frame variables. One can also write the operators which can act on expressions involving both
moving frame and spinor moving frame variables as if the latter were independent. These are
\bea\label{DIJ==}
D^{IJ}=\frac 12 v_\alpha^p\gamma^{IJ}_{pq}\frac {\partial } {\partial v_\alpha^p} + u_\mu^I \frac {\partial } {\partial u_\mu^J} - u_\mu^J \frac {\partial } {\partial u_\mu^I}\; , \qquad  \\ \label{DI==}D^{I}=\frac 12 v_\alpha^p\gamma^{I}_{pq}\frac {\partial } {\partial v_\alpha^p}+ u_\mu^0 \frac {\partial } {\partial u_\mu^I} + u_\mu^I \frac {\partial } {\partial u_\mu^0} \; . \qquad
 \eea

\subsection{From spinor moving frame to constrained helicity spinor. Prequel on SO(16) symmetry of twistor action. }
\label{sec:lambda=vS}

A rigorous development of twistor approach to
D$0$--brane  begins by introducing a new set of
constrained bosonic spinors $\lambda_\alpha{}^q$ related to the spinor moving frame matrix $v_\alpha{}^q$ by SO(16) rotation
\bea\label{l=vS}
\lambda_\alpha{}^p= v_\alpha{}^q S^{qp} , \qquad  SS^T=I_{16\times 16} \;  \qquad \\
\label{v=SL}\Rightarrow  \qquad v_\alpha{}^q =S^{qp}\lambda_\alpha{}^p\; . \qquad
\eea
These new objects with 165 degrees of freedom (vs 45 d.o.f-s in  $ v_\alpha{}^q $ summing 'physical' and pure gauge),  which we will call helicity spinors, obey the constraints\footnote{Hence, the name of constrained helicity spinors.  }
\begin{eqnarray}\label{u0s=ll}
u_\mu^{0} \sigma^\mu_{\alpha\beta}=\lambda_\alpha{}^q \lambda_\beta{}^q \; ,  \qquad
\nonumber \\  u_\mu^{I} \sigma^\mu_{\alpha\beta}=\lambda_\alpha{}^q (S^T\gamma^IS)_{qp}\lambda_\beta{}^p \; , \qquad
\\ \label{ltsl=} \lambda_{\alpha}^q \tilde{\sigma}{}_{\mu}^{\alpha\beta}\lambda_{\beta}^p= u_\mu^{0} \delta_{qp}+u_\mu^{I} (S^T\gamma^IS)_{qp}\; ,  \qquad \\ \nonumber   I=1,\ldots,9\; . \qquad
\end{eqnarray}
The matrices  $(S^T\gamma^IS)_{qp}$ carrying indices of $SO(16)$ are not constant but are symmetric and obey the same  Clifford algebra as  the constant $\gamma^I_{qp}$ do.
Notice that the upper and lower vector indices of $SO(16)$ group are transformed in the same way so that we are free in choosing their position and use this to make equations lighter and their structure more transparent.

Notice that \eqref{u0s=ll} as well as the contraction of
\eqref{ltsl=}, imply that $u_\mu^{0}$ is a composite of the constrained helicity spinors so that the spinor moving frame action  for D$0$--brane from  Eq. \eqref{eq:LD0} can be equivalently written in the following Ferber-Schirafuji-like form

\begin{equation}\label{eq:LD0=}
 {S}^{\text{D}0}=\, \int\limits_{{\cal W}^1} {\cal L}_{1}^{\text{D}0} = \frac m {16}\, \int\limits_{{\cal W}^1}\lambda_{\alpha}^q \tilde{\sigma}{}_{\mu}^{\alpha\beta}\lambda_{\beta}^q\,  \Pi^\mu -im \, \int\limits_{{\cal W}^1}\left(\text{d} \theta^{\alpha 1}\theta_{\alpha}^{2} - \theta^{\alpha 1}\text{d}\theta_{\alpha}^{2} \right)\; ,
\end{equation}
where  $\lambda_{\alpha}^q$ variables are constrained by \eqref{u0s=ll} and \eqref{ltsl=}. This action is clearly invariant under the $SO(16)$  gauge symmetry which acts on the Latin indices  ($p$) of
$\lambda_{\alpha}^p $
and on the matrix $S^{pq}$, involved in the constraints \eqref{u0s=ll} and \eqref{ltsl=}, by multiplication from the right. The $SO(9)$ gauge symmetry of the original form of the spinor moving frame action acts now on moving frame vectors $u_\mu^I$, entering the constraints
\eqref{u0s=ll} and \eqref{ltsl=},
and also on the matrix
$S^{pq}$ by multiplication from the left.

Thus our dynamical system, as described by the action
\eqref{eq:LD0=} and constraints \eqref{u0s=ll} and \eqref{ltsl=}, is invariant under the $SO(9)\times SO(16)$ gauge symmetry. The $SO(16)$ valued $S^{pq}(\tau)$ is clearly the Stueckelberg field for $SO(16)$ gauge symmetry which can be used to gauge $S^{pq}(\tau)$ to unity. Nevertheless, as we will see below, to maintain this field and to use constrained helicity spinors   $\lambda_{\alpha}{}^q$ instead of the original spinor moving frame variables $v_{\alpha}{}^q$  is very convenient for developing and for future using the twistor approach to D$0$--brane. This allows us to avoid the use of covariant momenta and covariant derivative in the formulation of the twistor  approach while maintaining the constraints    \eqref{u0s=ll} and \eqref{ltsl=} as strong equalities.
To explain why this is  possible,
in Appendix \ref{Dlambda} we introduce the covariant derivatives with respect to helicity spinors.

\subsection{Constrained supertwistor approach to D$0$--brane. I. Action and SO(16) symmetry }

\label{sec:YM-SO16}

 As it was in the case of
the massless $D=4$ superparticle, we would like to pass to a formulation
in terms of twistor and {{supertwistor}}  variables. In our $D=10$ case these supertwistor variables will be multiple and constrained.

To this end, we use \eqref{u0s=ll} and the Leibnitz rule to  present the action \p{eq:LD0=}
in the following form
\begin{eqnarray} \label{D0-1}
S^{\text{D}0} &=& m\int\limits_{\mathcal{W}^1}
\left (  \text{d} {\tilde x}^{\alpha\beta}
\cdot\lambda_\alpha{}^p \lambda_\beta{}^p -
i \text{d} \theta^{ \alpha 1} \cdot \theta^{ \beta 1}
\lambda_\alpha{}^p \lambda_\beta{}^p
-
i \text{d} \theta_{\alpha }^2 \,  \, \theta_{\beta}^{2}
\lambda_p{}^\alpha \lambda_p{}^\beta
- i \text{d} \theta^{ \alpha 1} \,  v_\alpha{}^p \,
\theta^2_p
+ i \theta^{ p 1} \text{d} \theta_\alpha^2
 \lambda_p{}^\alpha \right ) \\ \nonumber
&=&m\int\limits_{\mathcal{W}^1}
\left (  \text{d}( {\tilde x}^{\alpha\beta}
 \lambda_\alpha{}^p) \lambda_\beta{}^p +
(-{\tilde x}^{\alpha\beta}
 \lambda_\beta{}^q +
 i (\theta^{ p 1} \theta^{ q 1} +
 \theta^2_p \theta^2_q + 2 \theta^{p 1}
 \theta_q^2)\lambda_p{}^\alpha ) \text{d}
 v_\alpha{}^q
 -i \text{d} (\theta^{p 1 } + \theta^2_p)
 \cdot (\theta^{p 1} + \theta^2_p) \right )
\end{eqnarray}
Here  $ \lambda_q{}^\alpha$  is inverse to the helicity spinor matrix $ \lambda_\alpha{}^q$,
\be
 \lambda_\beta{}^p \lambda_p{}^\alpha = \delta{}_\beta{}^\alpha\; , \qquad
\lambda_p{}^\alpha \lambda_\alpha{}^q = \delta_{pq} \; , \ee
and we have also defined
\begin{equation}
\tilde{x}^{\alpha \beta} =
\frac{1}{16} x^\mu  {\tilde{\sigma}}_\mu\, ^{\alpha\beta}\; , \qquad
\theta^{q 1}= \theta^{\alpha 1}\lambda_\alpha^q\; , \qquad  \theta_q^{2}= \theta_\alpha^{2}\lambda_q^\alpha\; .
\end{equation}
Further, let us notice that
 (as can be seen from Eq. \eqref{ldl=}
in Appendix \ref{Dlambda}) the helicity spinors
satisfy in $\lambda_p{}^\alpha \text{d} \lambda_\alpha{}^p=0$.
Therefore,  the action
\eqref{D0-1} can be  written in the following twistorial form
\begin{equation} \label{twistoraction}
S^{\text{D}0}:= \int\limits_{\mathcal{W}^1} {\cal L}_1^{\text{D}0} = m\int\limits_{\mathcal{W}^1}
\left (
\text{d} \mu^{\alpha p} \cdot \lambda_\alpha{}^p
- \mu^{\alpha p} \cdot \text{d} \lambda_\alpha{}^p
- i \text{d} \eta^p \cdot \eta^p
\right ) =
m\int\limits_{\mathcal{W}^1}\text {d} Y^{M p}
\Omega_{MN} Y^{N p}\; ,
\end{equation}
as the difference between this functional and  \eqref{D0-1} vanishes, $
-\frac i {16}\theta^{p 1} \theta^2_p \lambda_q{}^\alpha \text{d} \lambda_\alpha{}^q
=0$.
In  \eqref{D0-1}
\bea \label{10dtwistor-2}
\mu^{\alpha p}=
{\tilde x}^{\alpha \beta} \lambda_\beta{}^p
- \frac{i}{2} (\theta^{ q 1} \theta^{ p 1} +
 \theta^2_q \theta^2_p + 2 \theta^{q 1}
 \theta_p^2)\lambda_q{}^\alpha+ \frac{i}{16} \theta^{q 1}\theta^2_q \lambda_p{}^\alpha, \nonumber \\
\qquad
\eta^p = \theta^{p 1} + \theta^2_p
\eea
and
\be\label{Omega}
\Omega_{MN} = \left(\begin{matrix}
0& \delta_\alpha{}^\beta  & 0\cr
 -\delta^\alpha{}_\beta & 0 & 0\cr
 0 & 0 & -i
\end{matrix}\right),
\ee
is an $OSp(32|1)$ invariant matrix.

In the second part of Eq. \eqref{twistoraction} we have introduced hexadecuplet of constrained  orthosymplectic supertwistors $Y^{Mp}$
\begin{equation} \label{10dtwistor-1}
Y^{M p} = (\mu^{\alpha p}, \lambda_\alpha{}^p, \eta^p)\; .
\end{equation}
The symplectic symmetry  $OSp(32|1)\times SO(16)$, which would be present in the action
\eqref{10dtwistor-2} with unconstrained supertwisors   \eqref{10dtwistor-1},
is broken down to $Spin(1,9)\times SO(16)$  by the constraints \eqref{u0s=ll} and \eqref{ltsl=} imposed on the helicity spinors $\lambda_\alpha^q$, which is used as the second $16\times 16$ bosonic block of the
$(32+1)\times 16$ supermatrix $Y^{Mp}$.

To see what at least the GL(16) subgroup $Sp(32)\in OSp(32|1)$ is about,   it is useful to note that gauge equivalent description of the system is given by the same action \eqref{twistoraction} but with $\mu^{\alpha p}$ of \eqref{10dtwistor-2} replaced by
\begin{equation} \label{10dtwistor-3}
\mu^{\alpha p}=
{\tilde X}^{\alpha \beta} \lambda_\beta{}^p
- \frac{i}{2} (\theta^{ q 1} \theta^{ p 1} +
 \theta^2_q \theta^2_p + 2 \theta^{q 1}
 \theta_p^2)\lambda_q{}^\alpha+ \frac{i}{16} \theta^{q 1}\theta^2_q \lambda_p{}^\alpha,
\qquad
\eta^p = \theta^{p 1} + \theta^2_p
\end{equation}
where ${\tilde X}^{\alpha \beta}={\tilde X}^{\beta\alpha }$ is an arbitrary symmetric spin-tensor. The enlarged superspace with  such 136 bosonic coordinates and some number $n$ of fermionic coordinates  clearly has $GL(16)$ automorphism symmetry of its natural supersymmetry algebra \footnote{Such tensorial superspaces were studied in relation with a description of massless higher spin theories  \cite{Bandos:1999qf,Vasiliev:2001zy,Plyushchay:2003gv,Bandos:2004nn,Bandos:2005mb,Sorokin:2017irs}, alternative to the unfolding approach of \cite{Vasiliev:1999ba}, as well as in relation with the so-called BPS preon concept of \cite{Bandos:2001pu}, \cite{Bandos:2002te,Bandos:2006xz,Bandos:2008um}. }.

The $SO(1,9)$ invariant decomposition of arbitrary $D=10$ spin-tensor reads
\be\label{tX=136} \tilde{X}^{\alpha \beta} =
\frac{1}{16} x^\mu  ({\tilde{\sigma}_\mu})^{\alpha\beta} +
\frac{1}{2\cdot 16\cdot 5!} y^{\mu_1\ldots \mu_5}  \tilde{\sigma}_{\mu_1\ldots \mu_5}^{\alpha\beta}\; , \qquad
\ee
The gauge equivalence of such a model in enlarged superspace with our original spinor moving frame formulation of D$0$--brane is suggested by the observation that  the action \eqref{twistoraction} is {\it formally} invariant under
\be\label{vmu=b5s5}
\delta \mu^{\alpha p} = b^{\mu_1\ldots \mu_5}   ({\tilde \sigma_{\mu_1\ldots \mu_5} })^{\alpha\beta}
\lambda_\beta{}^p\ee
(due to the constraints on the helicity spinors, see Eqs. \eqref{u0s=ll},  \eqref{ltsl=} and \eqref{ldl=} in the Appendix \ref{Dlambda}).
 The gauge symmetry \eqref{vmu=b5s5} allows to gauge away $126$ additional bosonic coordinate described by $D=10$ $5$-th rank selfdual antisymmetric tensor,
$y^{\mu_1\ldots \mu_5}=0$, thus reducing \eqref{tX=136} to the original form \eqref{10dtwistor-2} of our massive $D=10$ generalization of the Ferber--Penrose incidence relation  \eqref{4dtw}.

Actually, to make the above arguments on gauge symmetry \eqref{vmu=b5s5} of the action \eqref{twistoraction} literally valid, we should first notice that Eqs. \eqref{10dtwistor-3} with \eqref{tX=136} provide the general solution of the constraint
\be \label{jpq}
{\cal J}^{pq} := \mu^{\alpha [p} \lambda_\alpha{}^{q]} - \frac{i}{2} \eta^p \eta^q =0.
\ee
Thus we can consider our dynamical system to be described by the action \eqref{twistoraction} with supertwistor variables \eqref{10dtwistor-1} restricted, besides \eqref{u0s=ll} and \eqref{ltsl=}, by an additional constraint \eqref{jpq} only. Such action does possess the gauge symmetry \eqref{vmu=b5s5} as well as $SO(16)$ symmetry. As we will see in a moment, the constraint \eqref{jpq} serves as generator of this latter symmetry in the Hamiltonian approach.

\subsection{Constrained supertwistor approach to D$0$--brane. II Hamiltonian {{mechanics}}}

\label{sec:0-Tw-Ham}

In order to perform the Hamiltonian analysis
of the action \eqref{twistoraction}, we shall
follow the standard procedure, applicable for systems with bosonic and fermionic degrees of freedom \cite{Casalbuoni:1976tz}.
In particular, we find that  the calculation of canonical momentum for fermionic variable,
$\pi_{(\eta)}^p:= \frac {\partial {\cal L}^{D0}_1} {\partial d\eta^p}  = -im \eta^p$, gives us the constraint
\be\label{dp:=}
d_{(\eta)}^p:=\pi_{(\eta)}^p + im \eta^p \approx 0
\ee
stating that the real Grassmann variable
$ \eta^p$ is the momentum for itself.
To be explicit in this statement, we introduce the Poisson brackets
{
\be
\{ \pi_{(\eta)}^p, \eta^q  \}_{_{P.B.}}= \{  \eta^q , \pi_{(\eta)}^p  \}_{_{P.B.}}= -
\delta^{pq}, \qquad  \{ \eta^p, \eta^q  \}_{_{P.B.}}=0, \qquad  \{ \pi_{(\eta)}^p,  \pi_{(\eta)}^q \}_{_{P.B.}}= 0, \qquad
\ee
}}
and find that our fermionic constraints \eqref{dp:=} obey the algebra
\be\label{dfdf=} \{d_{(\eta)}^p, d_{(\eta)}^q \}_{P.B.}=-2im \delta^{pq}\ee
and, hence, are of the second class in classification of Dirac \cite{Dirac:1963}.

To treat these constraints as strong equalities, i.e. equalities which can be used before calculating all the Poisson brackets \cite{Dirac:1963}, we introduce Dirac brackets, defined as
\be\label{DB-G}
{}[...,...\}_{_{DB}}= [...,...\}_{_{PB}}- \frac{i}{2m} [..., d_{(\eta)}^p \}_{_{PB}} [d_{(\eta)}^p,...\}_{_{PB}}\; .
\ee
With respect to these Dirac brackets our real fermionic coordinate is self-conjugate,
\be\label{DB=f}
\{ \eta^p, \eta^q  \}_{_{D.B.}}=-\frac{i}{2m}
\delta^{pq},
\ee
which is the exact expression of the statement that
the fermionic coordinate function plays the role of its own momentum.

To analyze the bosonic sector of the dynamical system described by supertwistor action \eqref{twistoraction}, it is better to perform first the integration by parts reducing the second term of the Lagrangian form to the first one,
\be
{\cal L}_1^{\text{D}0} \mapsto  - m
\left (
 2\mu^{\alpha p} \cdot \text{d} \lambda_\alpha{}^p
+i \text{d} \eta^p \cdot \eta^p
\right )
\ee
Then the canonical momentum conjugate to $\lambda_\alpha{}^{p}$ is equal to $-2m\mu^{\alpha p}$, while momentum conjugate to
$\mu^{\alpha p}$ vanishes. These two relations are clearly the pair of the resolved second class constraints which can be used as string relations reducing the phase space after we pass to Dirac brackets
\be\label{DB=b}
[\mu^{\alpha p}, \lambda_\beta{}^{q}]_{D.B.}=
\frac{1}{2m}
\delta^\alpha_\beta \delta^{pq}\, . \qquad
\ee
The same results can be obtained by calculating the canonical momenta for bosonic variables directly from the Lagrangian form in \eqref{twistoraction} and then passing to Dirac brackets for the resulting second class constraints.

Of course, the  helicity spinor matrix
$\lambda_{\alpha}^p$  which is now a component of hexadecuplet of  $OSp(32|1)$ supertwisors \eqref{10dtwistor-1}, is strongly constrained by \eqref{u0s=ll}, \eqref{ltsl=}, which makes the  rigorous  development of Hamiltonian analysis quite  complicated. However, there exists a shortcut based on using the so-called covariant momenta for $\lambda_{\alpha}^p$ (classical counterparts of the covariant derivatives presented in Eqs.  \eqref{cDpq=} and \eqref{cDI=} of Appendix \ref{Dlambda}) instead of canonical momenta. This allows to treat the constraints
 \eqref{u0s=ll} and \eqref{ltsl=} as string relations without explicit search for corresponding Dirac brackets (see \cite{Bandos:2025pxv} and refs. therein).
% This simplifies the procedure essentially.
But moreover, studying the quantization in the frame of supertwistor formulation   \eqref{twistoraction},
 we actually do not need to specify
in detail this procedure, as the only additional constraint we will need to use in the quantization, Eq. \eqref{jpq}, contains the expression for one of such covariant momenta (the classical counterpart of ${\cal D}^{pq}$ in \eqref{fDqp} and  \eqref{cDpq=}) as its bosonic part and, thus, is well defined  when the constraints \eqref{u0s=ll}, \eqref{ltsl=} are treated as strong relations.

One can check that the constraints
\p{jpq}
form an algebra
\be
{} [{\cal J}^{pq}, {\cal J}^{rs} ]_{D.B.} = \frac{1}{2m}(\delta^{p[s} {\cal J}^{r] q}
- \delta^{q[s} {\cal J}^{r] p} )
\ee
under the Dirac brackets \eqref{DB=b} and \eqref{DB=f}.
This means, that these constraints
 are of the first class and generate
 $120$ parametric $Spin(16)$
 gauge symmetry.

\section{Quantization}
\label{quantization}

\subsection{Quantum D$0$--brane in constrained supertwistor approach} \label{QunatumD0-tw}

The quantization of the supertwisor formulation of the D$0$--brane (Dirichlet superparticle) which we describe in this (sub)section is quite similar to that for the quantization of
$D=11$ massless superparticle
sketched in
 \cite{Bandos:2006nr}
 (see
\cite{Green:1999by} for  quantization of
the massless  $D=11$ superparticle in the light-cone gauge, \cite{Bandos:2007wm} for similar quantization in spinor moving frame formalism and \cite{Berkovits:2002uc} for quantization
in terms of pure spinor variables).
This could have been expected,
taking into account the well known correspondence  between massless 11D   and massive 10D fields
(see e.g. sec. 5.3 of \cite{Green:1987sp} and also sec. 3 of \cite{Bandos:2025pxv}).

According to Dirac prescription, we should replace the basic Dirac brackets \eqref{DB=b} and \eqref{DB=f} by the commutator and anticommutaror,

\bea \label{comu}
[\hat{\mu}{}^{\alpha p}, \lambda_\beta{}^{q}]=
\frac{i}{2m}
\delta^\alpha_\beta \delta^{pq}, \qquad \\ \label{acomu}
\{ \hat \eta^p, \hat \eta^q  \}=\frac{1}{2m}
\delta^{pq},\qquad
\eea
One immediately observes that  fermionic operators $\hat \eta^p$ obey the 16 dimensional Clifford algebra. Hence they can be represented by SO(16)-invariant counterparts of Dirac matrices,
{{
\be\label{heta=Gamma} \hat{\eta}^q \;\mapsto \; \frac{1}{2\sqrt{m}}\,(\Gamma_q)_{\mathfrak A}{}^{\mathfrak B}
\ee
provided the quantum state vector of our dynamical system, $\Xi$ carries the $256$-valued index ${\mathfrak A}$ of the Dirac spinor of $SO(16)$,
\be\label{hetaXi=}
 \hat{\eta}^q \, \Xi_{\mathfrak A}=\, \frac 1{2\sqrt{m}}\, (\Gamma_q)_{\mathfrak A}{}^{\mathfrak B}  \Xi_{\mathfrak B}\; .
\ee
}}

Then, using the coordinate representation for the bosonic commutation relation
\eqref{comu},
\be
\hat{\mu}^{\alpha q} = \frac i {2m} \, \frac \partial {\partial \lambda_\alpha^q} , \qquad \hat{ \lambda}_\alpha^q =\lambda_\alpha^q\, , \ee we find that the first class constraints \eqref{jpq} imposed on the state vector reads
\be \label{10dqc}
\left ( \delta_{\mathfrak A}{}^{\mathfrak B} \left ( \lambda_{\alpha}{}^q \frac{\partial}{\partial \lambda_{\alpha p}} - \lambda_{\alpha}{}^p \frac{\partial}{\partial \lambda_{\alpha q}} \right )
 - \frac 1 2  \Gamma^{qp}{}_{\mathfrak A}{}^{\mathfrak B} \right ) \Xi _{\mathfrak B}(\lambda) =0\; .
\ee
It implies that the state vector transforms as a Majorana  spinor under $SO(16)$ symmetry acting also on the constrained helicity spinors. These carry degrees of freedom of the massive momentum $p_\mu =m u^0_\mu =\frac 1 {16} \, \lambda^q\tilde{\sigma}{}_\mu\lambda^q$ as well as an information on the polarization of different  massive particle states (see below).

{{
Now let us introduce the representation of
$SO(16)$ gamma matrices
\be\label{Gamma=si-si}  (\Gamma_q)_{\mathfrak A}{}^{\mathfrak B} = \left(\begin{matrix}0 & (\sigma_q)_{{\cal A}\tilde{\cal B}} \cr  \cr (\tilde{\sigma}_q)^{\tilde{\cal A}{\cal B}} & 0\end{matrix} \right)
\ee
in terms of $SO(16)$ counterparts of Pauli matrices carrying indices of different Majorana--Weyl spinor representations of $SO(16)$ and obeying
\be
(\sigma_q)_{{\cal A}\tilde{\cal C}}  (\tilde{\sigma}_p)^{\tilde{\cal C}{\cal B}}  + (\sigma_p)_{{\cal A}\tilde{\cal C}}  (\tilde{\sigma}_q)^{\tilde{\cal C}{\cal B}} = \delta_{pq}\delta_{{\cal A}}{}^{{\cal B}} \; , \qquad (\tilde{\sigma}_p)^{\tilde{\cal A }{\cal C}}  (\sigma_q)_{{\cal C}\tilde{\cal B}}   + (\tilde{\sigma}_q)^{\tilde{\cal A }{\cal C}}  (\sigma_p)_{{\cal C}\tilde{\cal B}}   = \delta_{pq}\delta^{\tilde{\cal A}}{}_{\tilde{\cal B}}\; ,
\ee
and also split  the quantum state vector of our dynamical system, $\Xi$ carries the $256$-valued index ${\mathfrak A}$ of the Dirac spinor of $SO(16)$ on two Majorana-Weyl spinors of $SO(16)$
\be\label{XifA=}
\Xi_{\mathfrak A}= \left(\begin{matrix}
\Phi_{\cal A} \cr  \Psi^{\tilde{\cal A}}  \end{matrix} \right)
\; . \qquad
\ee
}}
The action of the fermionic operator
$\hat{\eta}^q$ on these Weyl spinor components of the quantum state vector field are thus given by

\be\label{2mPhi=}
 2\sqrt{m}\,\hat{\eta}^q \,  \Phi_{\cal A} =   (\sigma_q)_{{\cal A}\tilde{\cal B}}  \Psi^{\tilde{\cal B}}
\; , \qquad \,  2\sqrt{m}\, \hat{\eta}^q \,  \Psi^{\tilde{\cal A}}=\,(\tilde{\sigma}_q)^{\tilde{\cal A}{\cal B}}  \Phi_{\cal B}\; .
\ee

Eq. \eqref{10dqc} can be split into two equations each for just one of two Majorana-Weyl component of the quantum state vector field,

\bea \label{10dqc=1}
\left ( \lambda_{\alpha}{}^q \frac{\partial}{\partial \lambda_{\alpha p}} - \lambda_{\alpha}{}^p \frac{\partial}{\partial \lambda_{\alpha q}} \right) \Phi _{\cal A} (\lambda)=
  \frac 1 2  \sigma^{qp}{}_{\cal A}{}^{\cal B}  \Phi _{\cal B} (\lambda)\; , \\
 \label{10dqc=2}
 \left ( \lambda_{\alpha}{}^q \frac{\partial}{\partial \lambda_{\alpha p}} - \lambda_{\alpha}{}^p \frac{\partial}{\partial \lambda_{\alpha q}} \right )  \Psi^{\tilde{\cal A}}(\lambda) =
  \frac 1 2  \tilde{\sigma}^{qp}{}^{\tilde{\cal A}}{}_{\tilde{\cal  B}}\Psi^{\tilde{\cal B}}(\lambda)\; .
\eea

The fermionic nature of $\hat{\eta}^q$ operator and Eq. \eqref{XifA=} suggests
to consider these two Majorana-Weyl spinor components of the state vector field to be of different statistics, e.g.
bosonic $\Phi_{\cal B}$ and fermionic $\Psi^{\tilde{\cal B}}$.
The suitable  $SO(9)$ covariant representation of this latter can be given
in terms of gamma--traceless 9d vector-spinor, while the former can be split onto symmetric traceless second rank tensor and antisymmetric third rank tensor

\bea
\Phi_{\cal B}&=& \left(\begin{matrix} h_{IJ}=h_{JI}\; ,\qquad h_{II}=0\; ,
\cr  a_{IJK}=a_{[IJK]}  \end{matrix} \right)\; , \qquad \nonumber \\ \nonumber \\  \Psi^{\tilde{\cal A}} &=& \left(\sqrt{\frac {2}{m}}\tilde{\psi}^I_q\; , \qquad  \tilde{\psi}^I_q (S^T\gamma^IS)_{qp}=0\right)\; . \qquad
\eea

The explicit $SO(9)$ invariant representation of $SO(16)$ Pauli matrices in these split notation can be found in \cite{Bandos:2007wm}. With it, the realization of the $SO(16)$ Clifford algebra on the above set of fields is given by

\bea\label{etah=}
&&\hat{\eta}^q\, h_{IJ} \, =\, \frac 1 m \,  (S^T\gamma^{(I}S)_{qp}
\psi^{J)}_p\; , \qquad \nonumber \\ \nonumber \\  && \hat{\eta}^q\, a_{IJK} \, = \frac 3 {2m} \, (S^T\gamma^{[IJ}S)_{qp}
\psi^{K]}_p\; , \qquad  \\ \nonumber \\  \label{etaPsi=}
&& \hat{\eta}^q\,
\psi^{I}_p = \frac 1 2 \, h_{IJ}(S^T\gamma^JS)_{qp} + \frac 1 {12}\, a_{JKL}\,  \left(S^T\gamma^{IJKL}S-6\delta^{I[J}S^T\gamma^{KL]}S\right){}_{qp}  \,\; , \qquad
\eea
({\it cf.} \cite{Green:1998by,Bandos:2006nr,Bandos:2007wm} where quantization of massless $D=11$ superparticle was considered).

The above set of fields with $SO(9)$ indices are related with
the solutions of linearized equations of the massive counterpart  of type IIA supergravity by
\be \label{10dsols}
h_{\mu \nu} (p) = u_\mu^I u_\nu^J
h_{IJ}(\lambda), \qquad
a_{\mu \nu \rho} (p) = u_\mu^I u_\nu^J u_\rho^K
a_{IJK}(\lambda)\; , \qquad
{\psi}_{\mu \alpha} (p) =
u^I_\mu \lambda_{\alpha q} \tilde{\psi}_{Iq}(\lambda).  \qquad
\ee
Here, let us recall, the moving frame vectors $ u_\mu^I$ are spacelike, orthogonal, normalized and related to the helicity spinors
$\lambda_{\alpha q}$ by the constraints \eqref{ltsl=} while
the momentum is timelike and expressed by
\be
p_\mu = mu^0_\mu = \frac m {16}  \, \lambda_{\alpha q} \tilde{\sigma}_\mu^{\alpha\beta}\lambda_{\beta q} \; . \ee
Fixing the $SO(16)$ gauge $S^{qp}=\delta^{qp}$, in which
$\lambda_\alpha^{q}=v_\alpha^{q}$,
Eqs. \eqref{10dsols} can be reduced to the ones obtained in \cite{Bandos:2025pxv}, where it was shown that quantum state spectrum of D$0$--brane was given by the massive counterpart of type IIA supergravity and its relation with the fields of linearized $D=11$ supergravity was considered \footnote{In \cite{Bandos:2025pxv} one can also find the explanation of why we prefer not to use the name ''massive type IIA supergravity'' for the massive counterpart of type IIA supergravity. }.
This conclusion about quantum state spectrum of D$0$--brane is thus confirmed by supertwistor quantization of the dynamical system which we have discussed above.

Our supertwistor approach to quantum D$0$--brane possesses invariance under the $SO(16)$ symmetry which implies imposing the Eqs.  \eqref{10dqc=1} on the bosonic fields and Eqs. \eqref{10dqc=2} on the fermionic fields. The r.h.s.-s of these equations can be specified either by using the explicit representation for $SO(16)$ Pauli matrices or from  antisymmetric
$[rq]$ part of the action by $\hat{\eta}^{r}$ on Eqs.  \eqref{etah=} and \eqref{etaPsi=}.  We do not need the explicit form of these equations for our discussion below.
Let us only note that equations for $h_{IJ}$ ($A_{IJK}$)  hidden in
\eqref{10dqc=1}  involves $a_{IJK}$ ($h_{IJ}$) while equation for fermionic field, \eqref{10dqc=1} does not involve other fields.

In the next section we will use another quantization scheme for our dynamical system. This will relate the results of supertwistor quantization as it is performed in this section with the spinor moving frame description of quantum D$0$--brane given in \cite{Bandos:2025pxv}.

%\bigskip

\subsection{Alternative quantization I. Constrained  on--shell superfields from supertwistor approach. }
\label{Quant=2}

Let us come back to the fermionic second class constraints
\eqref{dp:=} which obey the algebra \eqref{dfdf=} on the Poisson brackets. In the coordinate representation
the quantum counterpart of these constraint, $\hat{d}_{(\eta)}^q$, is proportional to
\be
D_q=\frac {\partial } {\partial \eta^q}-m\eta^q
\ee
($\hat{d}_{(\eta)}^q=-iD_q$) which obeys
\be\label{DpDq=-2m}
\{ D_q, D_p\}  = -2m \delta_{qp} \; .
\ee
Clearly we cannot impose these constraints on the state vector in the sense that state vector vanishes under their action. However, what we can is to {\it represent the state vector by a set of superfields} in such a way that the result of the action of
$ D_q$ on one of such superfields is expressed through the other superfields in a self-consistent
manner.

Such a way  of quantization was briefly discussed in  \cite{Bandos:2017eof} in the context of $D=11$ supergravity for which it resulted in the set of superfield equations first proposed in \cite{Galperin:1992pz}. These are quite similar
to the  system of superfield equations for massive counterpart of type IIA supergravity appearing in quantum state spectrum of D$0$--brane \cite{Bandos:2025pxv} which we will describe below.

The suitable set of superfields to describe the quantum state spectrum of the  D$0$--brane in the frame of our constrained supertwistor approach is suggested by the study in the previous section (as the operator $\hat{\eta}^q$ there obey
the Clifford-like algebra similar to  \eqref{DpDq=-2m}). It contains the bosonic symmetric traceless SO(9) tensor  superfield $H_{IJ}(\lambda, \eta)$, antisymmetric third rank  tensor superfield $A_{IJK}(\lambda, \eta)$ and the fermionic superfield with $SO(9)$ vector and $SO(16)$ vector indices, $\Psi^I_q(\lambda, \eta)$  obeying $\Psi^I_q(S^T\gamma^IS)_{qp}=0$,

\be \label{superf=const}
\begin{cases}
H_{IJ}(\lambda, \eta)= H_{JI}(\lambda, \eta)\; , \qquad H_{II}(\lambda, \eta)=0
\cr
A_{IJK}(\lambda, \eta)=A_{[IJK]}(\lambda, \eta)\; ,
\cr
\Psi^I_q(\lambda, \eta)\; , \qquad
\Psi^I_q(S^T\gamma^IS)_{qp}=0\; .
\end{cases}
\ee

The superfield equations representing the fermionic constraint action on the state vectors are
\bea\label{DqH=}
D_q\, H_{IJ} & =& 2i\, (S^T\gamma^{(I}S)_{qp}
\Psi^{J)}_p\; , \qquad \\ \label{DqAIJK=} D_q\,A_{IJK} & = & 3i \, (S^T\gamma^{[IJ}S)_{qp}
\Psi^{K]}_p\; , \qquad  \\ \label{DqPsi=}
D_q
\Psi^{I}_p &=&im\, H_{IJ}(S^T\gamma^JS)_{qp} + \frac {im} {3!}\, A_{JKL}\,  \left(S^T\gamma^{IJKL}S-6\delta^{I[J}S^T\gamma^{KL]}S\right){}_{qp}  \,\; . \qquad
\eea

To deal with the constraints \eqref{jpq} in this approach we should first use the fermionic constraint \eqref{dp:=} to write it as
\be \label{jpq==}
{\cal J}^{pq} = \mu^{\alpha [p} \lambda_\alpha{}^{q]} + \frac{1}{2m} \eta^{[q} \pi_{(\eta)}^{p]}
\ee
which is the true first class constraint in the presence of second class constraint
\eqref{dp:=}.

The quantum version of $2mi{\cal J}^{pq}=-2mi{\cal J}^{qp} $ is given by differential operator
\be \label{fDqp}
{\mathfrak D}^{qp} =  \lambda_\alpha{}^{[q}
\frac {\partial}  {\partial \lambda_\alpha{}^{p]}}
+  \eta^{[q} \frac {\partial}  {\partial \eta^{p]}}
\ee
which generates $SO(16)$ transformations of
$\lambda_\alpha{}^p$, $\eta^p$ as well as of $S^{pq}=\lambda_\alpha{}^pv_q^\alpha$ and $D_q$:

\bea\label{fDqp1}
{\mathfrak D}^{qp} \lambda_\alpha{}^r =
\lambda_\alpha{}^{[q}\delta^{p]r}\; , \qquad {\mathfrak D}^{qp} \eta^r =
\eta^{[q}\delta^{p]r}\; , \qquad
\\ \label{fDqpS=}
{\mathfrak D}^{qp} S^{rs} =
S^{r[q}\delta^{p]s}\; , \qquad \\ \label{fDqpDr=}
{}[{\mathfrak D}^{qp}, D_r]= D_{[q}\delta_{p]r}\; .  \qquad
\eea

Our multicomponent state vector superfield should be restricted by imposing in some manner the quantum constraint    generating $SO(16)$ symmetry. The consistency with \eqref{DqH=}, \eqref{DqPsi=} and \eqref{fDqpDr=} does not allow
to consider all the component superfields describing our  state vector to be in kernel of the differential operator ${\cal J}^{pq} =-\frac i {2m}{\mathfrak D}^{qp}$.
To be precise, this is possible for the  bosonic superfield, but not for fermionic one, so that
we impose
\bea\label{fDpq=b}
{\mathfrak D}^{qp} H^{IJ}=0\; , \qquad {\mathfrak D}^{qp} A_{IJK}=0\; , \qquad \\
\label{fDpq=f} {\mathfrak D}^{qp} \Psi^{I}_r=  \Psi^{I}_{[q}\delta_{p]r}\; .  \qquad
\eea

Actually, this is logical since  the fermionic field carries $SO(16)$ vector index $q$ while the bosonic superfields carry  SO(9) vector indices only. The technical reason is the nontrivial action on the fermionic derivative $D_p$ and on $S^{pq}$ matrix variable, \eqref{fDqpDr=} and \eqref{fDqpS=}.
To resume, the constraints \eqref{fDpq=b}  and  \eqref{fDpq=f}  imply the invariance of bosonic superfields and covariance of the fermionic superfield
under $SO(16)$ symmetry which also acts on bosonic and fermionic components of constrained supertwistor coordinates.

It can be shown that the field content of the set of superfields \eqref{superf=const} is given by the set of their leading components,
\be\label{h=H0}
h^{IJ}= H^{IJ}\vert_{\eta^q=0}\; , \qquad
a_{IJK}= A_{IJK}\vert_{\eta^q=0}\; , \qquad \psi^{I}_q= \Psi^{I}_q\vert_{\eta^q=0}\; , \qquad
\ee
while the higher components of each of these superfields are expressed in terms of leading components of the other ones. Indeed,
the next to leading component of fermionic superfield,
$D_p\Psi^I_q\vert_{\eta =0}$,  is expressed through the leading components of the bosonic superfields by  leading component of Eq. \eqref{DqPsi=}, while the next to leading   components of the bosonic superfields are expressed in terms of leading component of the fermionic superfield by using Eqs. \eqref{DqH=} and \eqref{DqAIJK=}. The independent component fields \eqref{h=H0}, in their turn, describe  the solutions \eqref{10dsols} of the linearized equations of massive analog of the type IIA supergravity multiplet \cite{Bandos:2025pxv}.

\subsection{Alternative quantization II. Analytic on shell superfields from constrained supertwistor approach. }
\label{Quant=3}

In this section we will show how the above constrained superfield description of
D$0$--brane quantum state vector can be related with the description by analytic  (chiral) superfield subject to a duality constraint obtained in \cite{Bandos:2025pxv}.

To this end we first have to introduce, following
\cite{Bandos:2017zap} and
\cite{Bandos:2025pxv}, the
complex null vector $U^I$, $I=1,...,9$, normalized on 2 in its contraction with its conjugate    $\bar{U}^I$, and seven real normalized vectors orthogonal to these and among themselves,
\begin{eqnarray}\label{UU=0=} && U_IU_I=0=  \bar{U}_I \bar{U}_I \; , \qquad U_I \bar{U}_I=2 \; , \qquad \nonumber \\ && U_IU_I{}^{\check{J}} =0=  \bar{U}_I U_I{}^{\check{J}} \; , \qquad  U_I{}^{\check{J}}U_I{}^{\check{K}}  =\delta^{\check{J}\check{K}}\; . \qquad \end{eqnarray}
This set of vectors form a basis in 9d Euclidean space and can be collected   in the $SO(9)$ matrix
\begin{eqnarray}
\label{UinSO9} &&
  U_I^{(J)}= \left(U_I{}^{\check{J}}, \frac 1 2 \left( U_I+ \bar{U}_I\right), \frac 1 {2i} \left( U_I- \bar{U}_I \right)\right) \; \in \; {SO}(9)
 \;  . \qquad
  \end{eqnarray}
The splitting in \eqref{UinSO9} is invariant under $SO(7)\times U(1)$ symmetry which allows to consider $U_I, \bar{U}_I$ and  $U_I{}^{\check{J}}$ as constrained homogeneous coordinates of the coset $\frac {SO(9)} {SO(7)\times U(1)}$.

The ''covariant harmonic derivatives'', this is to say such differential operators acting on the vectors $U_I, \bar{U}_I, U_I{}^{\check{J}}$ which preserve constraints
\eqref{UU=0=}, are

\bea \label{bDchI=}
{{\bb D}}^{\check{I}}= \frac 1 2 \, U_J\, \frac {\partial } {\partial U_J^{\check{I}}}- U_J^{\check{I}}\, \frac {\partial } {\partial \bar{U}_J}\; , \qquad
\bar{ {\bb D}}^{\check{I}}= \frac 1 2 \, \bar{U_J}\, \frac {\partial } {\partial U_J^{\check{I}}}- U_J^{\check{I}}\, \frac {\partial } {\partial {U}_J}\; , \qquad \\ \label{bDchIchJ=}
 \bar{{\bb D}}^{\check{I}\check{J}}= \frac 1 2 \, U_K^{\check{I}}\, \frac {\partial } {\partial U_K^{\check{J}}}- U_K^{\check{J}}\, \frac {\partial } {\partial U_K^{\check{I}}}\; , \qquad  {\bb D}^{[0]}= U_J\, \frac {\partial } {\partial {U}_J}-\bar{U}_J\, \frac {\partial } {\partial \bar{U}_J}\; . \qquad
\eea
Of these, \eqref{bDchI=} are vector fields tangential to the coset $\frac {SO(9)} {SO(7)\times U(1)}$ while \eqref{bDchIchJ=} are in $SO(7)$ and $ U(1)$ directions.
Their commutators  provide us with a manifestly  $SO(7)\times U(1)$ covariant representation of $so(9)$ algebra.

Next, we have to introduce the complex $16\times 8$ complex matrix variables
${\rm w}_q^A= (\bar{{\rm w}}_{qA})^*$ providing the bridges between
the $16$ dimensional vector representation of $SO(16)$ and fundamental (anti-fundamental) representation  of $SU(8)$ group ({\it cf.}
\cite{Bandos:2025pxv}).
The complex rectangular matrix $\bar{{\rm w}}_{qA}=({{\rm w}}_{q}^{A})^*$
is related to the block $\bar{w}_{p\tilde{A}}=({{w}}_{q}^{\tilde{A}})^*$ of a complex representation of a $Spin(9)$ valued matrix,  $(\bar{w}_{p\tilde{A}}, \, w_p^{\tilde{A}})\in Spin(9)$ with $SO(7)$ spinor index ${\tilde{A}}=1,...,8$ and $SO(9)$ spinor index $p$, by:
\begin{itemize}
\item  left multiplication by $SO(16)$  matrix $S^{qp}$  (see \eqref{l=vS} and \eqref{S=vl}),  so that $q$ is converted to the $SO(16)$ index $p$, and
\item  action  of $SU(8)$ group  in its anti-fundamental representation to obtain
 $\bar{{\rm w}}_{qA}=({{\rm w}}_{q}^A)^*$.
\end{itemize}

The action of anti-fundamental representation of $SU(8)$ on the indices of $\delta_{\tilde{A}\tilde{B}}$, which plays the role of SO(7) charge conjugation matrix, gives a non-constant complex symmetric matrix ${\cal U}_{AB}$ while
the action on that by fundamental representation gives the complex conjugate matrix
$\bar{{\cal U}}^{AB}$. Clearly this complex matrix  ${\cal U}_{AB}$ is symmetric and unitary, i.e. it obeys
\be
{\cal U}_{AC}\bar{{\cal U}}^{CB}=\delta_A{}^B\; , \qquad {\cal U}_{AB}={\cal U}_{BA} = (\bar{{\cal U}}^{AB})^*\; . \qquad
\ee
See  Appendix \ref{app:rmW} and  \cite{Bandos:2025pxv} for more details on  representations of $SU(8)$ and $SO(7)$.

When  Spin(9) valued matrix  $(\bar{{w}}_{q\tilde{A}}, \, {w}_q^{\tilde{A}})$ is the double covering of the SO(9) frame \eqref{UinSO9}, the   bridge variables $(\bar{{\rm w}}_{qA}, \, {\rm w}_q^{{A}})$ are related with this SO(9) vector frame by Eqs. \eqref{wgIw=UcU} and \eqref{UIgI=}  presented in Appendix \ref{app:rmW}. The true constraints which restrict these variables
 are
\begin{eqnarray}
\label{brwrw=I}
&& \bar{{\rm w}}_{qA}\bar{{\rm w}}_{qB}=0 ={\rm w}_q^A {\rm w}_q^B\; , \qquad \bar{{\rm w}}_{qA} {\rm w}_q^B=\delta_A{}^B\; , \qquad \bar{{\rm w}}_{qA} {\rm w}_p^A +  {\rm w}_q^A \bar{{\rm w}}_{pA} = \delta_{qp}  \; . \qquad
\end{eqnarray}
These imply that they provide a complex representation for $SO(16)$ matrix.

Using these variables and complex null vector from the  SO(9) frame \eqref{UinSO9} we can construct from  \eqref{superf=const} complex  bosonic and fermionic superfields inert under SO(9)

\be\label{Phi=HUU}
\Phi = H^{IJ}U_IU_J\; , \qquad \Psi^{A(U)}= \Psi^I_q U_I {\rm w}_q^A\; , \qquad \bar{\Psi}_A^{(U)}= \Psi^I_q U_I \bar{{\rm w}}_{qA}\; , \qquad etc. ,
\ee
as well as mutually complex conjugate octuplets of the fermionic covariant derivatives

\be\label{bDA=DwA}
{D}{}^A= {\rm w}_q^A D_q \; , \qquad \bar{D}{}_A= \bar{{\rm w}}_{qA} D_q\; . \qquad
\ee
It is not difficult to find that $D_q\Phi = 4i  \bar{{\rm w}}_{qA}{\cal U}^{AB}\bar{\Psi}_B^{(U)}$. Contracting this with $ {\rm w}_q^A$ we find that the superfield $\Phi$ is chiral,
\be\label{bDAPhi=0}
\bar{D}{}_A\Phi=0\; ,
\ee
while contracting with $ {\rm w}_q^A$ we find
$D^A \Phi= 4i {\cal U}^{AB}\bar{\Psi}_B^{(U)}$ which implies the dependence of the fermionic superfield, $\bar{\Psi}_B^{(U)}=-\frac i 4 \bar{{\cal U}}_{BA}D^A \Phi$.

The superfield $\Phi$ also obeys
\be\label{bD4Phi=}
D^AD^BD^CD^D \Phi= \frac 1 {4!} \epsilon^{ABCDEFGH }\bar{D}{}_E\bar{D}{}_F\bar{D}{}_G\bar{D}{}_H(\Phi)^*\; .
\ee
To prove this is more involving;
however the direct calculations are actually not necessary to be convinced that it holds (at least up to the sign).    Indeed, it is sufficient to notice that: \begin{itemize}
    \item  the set of superfields \eqref{superf=const} obeying \eqref{DqH=}--\eqref{DqPsi=} describes the massive counterpart of type IIA supergravity multiplet (see \cite{Bandos:2025pxv}); \item as it was shown in \cite{Bandos:2025pxv}, the same multiplet is described by the chiral superfield obeying
\eqref{bDAPhi=0} and \eqref{bD4Phi=}; \item  all the set of superfields
\eqref{superf=const} can be reproduced from the analytical  superfield
$\Phi$ in \eqref{Phi=HUU} with the use of covariant harmonic derivative of $SO(9)$ internal frame \eqref{bDchI=}, fermionic derivative and Eqs. \eqref{DqH=}--\eqref{DqPsi=}.
\end{itemize}
(Only sign on \eqref{bD4Phi=} requires an explicit calculation to be fixed as it can be changed by replacing $\Phi$ by $i\Phi$).

The last item above probably needs some evidences. To have these, one can easily check that,  for instance,
\bea
&& \bar{ {\bb D}}^{\check{I}}\Phi =- 2 U_JH^{JK}U_K^{\check{I}}\; , \qquad \nonumber \\ && \bar{ {\bb D}}^{\check{I}}\bar{ {\bb D}}^{\check{J}}\Phi = 2 U_K^{\check{I}}H^{KL}U_L^{\check{J}}-U_K^{\check{I}'}H^{KL}U_L^{\check{I'}}\delta^{\check{I}\check{J}}=  {\bb D}^{\check{I}}{\bb D}^{\check{J}}\bar{\Phi}\; , \qquad
\eea
which implies
\bea
H^{IJ}=\frac 1 4 \bar{U}_I\bar{U}_J \Phi - \frac 1 2 \bar{U}_{(I}{U}_{J)}^{\check{K}} {\bb D}^{\check{K}}\Phi  + {U}_I^{\check{K}}{U}_J^{\check{L}}\left( \frac 1 4 {\bb D}^{\check{K}}{\bb D}^{\check{L}}\Phi   -  \frac 1 {36} {\bb D}^{\check{I}'}{\bb D}^{\check{I}'}\Phi \delta^{^{\check{K}\check{L}}} \right) +  \frac 1 {36} \bar{U}_{(I}{U}_{J)}{\bb D}^{\check{I}'}{\bb D}^{\check{I}'}\Phi  +c.c.
\eea

The action of the covariant derivatives \eqref{bDchI=}  on the ${\rm w}_q^A, \bar{{\rm w}}_{qA}$  variables is described in Eqs. \eqref{bbDIw=}
of Appendix \ref{app:rmW}. These can be reproduced using the $SO(16)$ rotations and $SU(8)$ transformations from their action on $Spin(9)$ variables which can be found in  \cite{Bandos:2017eof}.

\bigskip

\section{Conclusions and discussion}
\label{Conclusion}

In this paper we have developed a constrained supertwistor approach to D$0$--brane in classical and quantum domain. This provides a ten dimensional generalization  of the two--twistor approach to $D=4$ massive particle and superparticle from \cite{Bette:2004ip,Fedoruk:2005ks,deAzcarraga:2005ky,deAzcarraga:2008ik}. However, the structure and symmetries in ten dimensional case are much {{richer}}.

Although the basis for {{this}} supertwistor approach has been  provided by spinor moving frame formulation of D$0$--brane from \cite{Bandos:2000tg,Bandos:2025pxv},  the basic bosonic spinor variables of the supertwistor approach are not identified with spinor moving frame matrix $v_\alpha{}^q \in Spin(9)$ but with a helicity spinor matrix $\lambda_\alpha{}^p$ which is obtained from $v_\alpha{}^q$ by $SO(16)$ transformation,
$\lambda_\alpha^p= v_\alpha^q S^{qp}$.
This allows to avoid the use of the so--called covariant momenta and covariant derivative formalisms  (described in Appendix \ref{Dlambda}) and to study the properties of the quantum state vector of D$0$--brane using with the standard partial derivatives with respect to  the bosonic spinors $\lambda_\alpha^p$ in a manner consistent with the constraints imposed on these latter.

The variables of the supertwistor approach to D$0$--branes are hexadecuplet of the constrained $OSp(32|1)$
supertwistors $Y^{Mp}=(\mu^{\alpha p}, \lambda_\alpha{}^p,\eta^p)$ subject to the constraint
generating the $SO(16)$ symmetry in the Hamiltonian approach. The index $p=1,...,16$ enumerating the supertwistors is transformed by vector representation of $SO(16)$. $OSp(32|1)$ symmetry is broken by the constraints imposed on the $\lambda_\alpha{}^p$ component of the supertwistor.
However, at least the $GL(16)$ subgroup of this
symmetry can be followed to the bosonic sector of the so called tensorial superspace which appears in the general solution of the $SO(16)$ constraints on supertwistor variables, which provides the generalization of the Ferber-Penrose incidence relations \cite{Ferber:1977qx}.

We quantize the supertwistor formulation of D$0$--brane by two schemes. In the first the quantum state vector is represented by 256 component  Majorana
spinor of $SO(16)$, the Majorana-Weyl ''constituents'' of which are taken to be with opposite statistics and
 (after gauge fixing of $SO(16)$ symmetry) can be represented   by fermionic gamma-traceless $SO(9)$ vector--spinor field, and by the set of  bosonic third rank antisymmetric and bosonic second rank symmetric traceless tensors of $SO(9)$.

The second of the quantization schemes describes the quantum  state vector by a set of superfields subject to some equations with fermionic covariant derivatives. The leading components of these constrained superfields can be associated with the above describe set of fields. These, in their turn,
can be used to write the solutions of the linearized spacetime equations of the fields of the massive counterpart of type IIA supergravity multiplet which thus describes the quantum state spectrum of D$0$--brane.

Using these constrained superfield approach to quantization, we have shown explicitly how the description of the D$0$--brane quantum state vector by analytical on-shell superfield obtained in \cite{Bandos:2025pxv} can be derived from the supertwistor approach.

The fact that the on-shell superfields are one-particle counterparts of the on-shell superamplitudes, which in the $D=4$ case were the basis of an impressive progress in calculation of tree and loop processes in maximally supersymmetric Yang--Mills and supergravity theories, gives us the reason to hope  that our results will be useful for development of the on-shell superamplitude approach in type IIA string theory, in its part involving D$0$-branes besides supergravity multiplets. The preliminary studies  in \cite{Bandos:2025pxv} indicated a problem hampering the way towards an analytic  (chiral) superamplitude formalism for such type processes. A possible way out might pass through the development of a superamplitude counterpart of the constrained on-shell superfield formalism obtained in sec. \ref{Quant=2} or of an amplitude counterpart {{of}} the $SO(16)$ spinor description of the quantum state vector of D$0$--brane which we have obtained in sec. \ref{QunatumD0-tw}.

Another possible application of our approach can be related with the study of (super)twistor structures in curved $D=10$ spacetimes on the line of $D=4$ two-twistor approach to studying Kerr--Newman black holes developed in recent \cite{Kim:2026yqo}.

We hope to address these problems in the near future.

\bigskip

\section*{Acknowledgments}
M.T. would like to thank the Department of Physics
of the National Tsing Hua University, Hsinchu, Taiwan, and the Department of Physics of
the National Sun Yat-sen University, Kaohsiung, Taiwan for their hospitality during the initial stage of the project.
The work of I.B.  has been supported in part by the MCI, AEI,
FEDER (UE) grant  PID2024-155685NB-C21  (“Gravity, Supergravity and Superstrings” (GRASS)) and, at initial stage of this project,  by the Basque Government grant IT-1628-22. The work of M.T. was supported by the Quantum Gravity Unit of the Okinawa Institute of Science and Technology Graduate University (OIST).

\appendix

\section{ Notation and Conventions}
\label{DC}
%In $D=4$ we use the following indices
%\begin{itemize}
%\item The Greek letters from the middle %of the alphabet correspond to vector %indices
%$\mu, \nu, \ldots=0, \ldots ,3$.

%\item The Greek letters from the %beginning of the alphabet and small %letters from the middle of the Latin %alphabet are $SL(2,C)$
%spinor indices
%$\alpha, \beta, \ldots =1,2$ and
%$\dot \alpha, \dot \beta, \ldots =1,2$

%\end{itemize}

In $D=10$ we use the following sets of indices
\begin{itemize}
\item The Greek symbols from the middle of the alphabet correspond to vector indices:
$\mu, \nu, \ldots=0, \ldots ,9$.

\item The Greek symbols from the beginning of the alphabet
$\alpha, \beta, \ldots =1, \ldots , 16$ denote the $SO(1,9)$ spinor indices

\item The small letters from the middle of Latin alphabet, $p,q, \ldots =1, \ldots , 16$
denote $SO(16)$ vector indices of some variables, in particular $\lambda_\alpha{}^q$, and spinor $SO(9)$ $(Spin(9))$ indices of other variables, like $v_\alpha{}^q$. We believe this will not produce any
confusion, and certainly lighten the notation. The orthogonal matrix $S^{qp}$ carries first index, $q$, of $Spin(9)$ and the second index, $p$, of $SO(16)$;  $S^{qp}=S^{-1\, pq}$.

\item
The capital Latin letters $I,J,K, \ldots =1,...,9$. denote $9$--vector indices which are transformed by $SO(9)$ group.

\item
The capital Latin letters with check symbol, $\check{I},\check{J},\check{K}, \ldots =1,...,7$. denote 7--vector indices which are transformed by $SO(7)$ group.

\end{itemize}
The generalized $D=10$ Pauli matrices
$\sigma^\mu_{\alpha\beta}=\sigma^\mu_{\beta\alpha}$ and  $\tilde{\sigma}{}^{\mu\alpha\beta}=\tilde{\sigma}{}^{\mu\beta\alpha}$ obey
\be
\sigma^\mu\tilde{\sigma}{}^\nu+ \sigma^\nu\tilde{\sigma}{}^\mu = \eta^{\mu\nu} {\bb I}_{16\times 16}\; , \qquad  \eta_{\mu \nu}= diag(1,-1,...,-1).
\ee
The $d=9$ Dirac matrices  $\gamma^I_{qp}=
\gamma^I_{pq}$   obey the Clifford algebra
\begin{eqnarray}\label{gIgJ=dIJ=A}
\gamma^I_{qp} =\gamma^I_{pq} \; ,  \qquad \gamma^I\gamma^J+\gamma^J\gamma^I=2\delta^{IJ} {\mathbb I}_{16\times 16} \; .  \qquad
\end{eqnarray}

The complex antisymmetric 8$\times 8$ matrices $\sigma^{\check{I}}_{AB}=-\sigma^{\check{I}}_{BA}$ and $\tilde{\sigma}{}^{\check{J}\; AB}=-\tilde{\sigma}{}^{\check{J}\; BA}=-(\sigma^{\check{I}}_{AB})^*$, carrying the indices of fundamental and anti-fundamental representations of $SU(8)$, are obtained from $SO(7)$ counterparts of Pauli matrices, imaginary antisymmetric  $\sigma^{\check{I}}_{\tilde{A}\tilde{B}}= -\sigma^{\check{I}}_{\tilde{B}\tilde{A}}$,  by $SU(8)$ transformations in the corresponding representation. They obey \eqref{sts=SU}
(see \cite{Bandos:2025pxv} for more details).

\section{Covariant derivatives of the constrained helicity spinor formalism and some comments}
\label{Dlambda}

In this Appendix we introduce  the counterparts of the covariant harmonic derivatives \eqref{DIJ=}  for the  constrained helicity spinors $\lambda_\alpha^q$ obeying \eqref{u0s=ll}, \eqref{ltsl=}.
 First step in this direction is to find the expression for admissible derivative of helicity spinor matrix, which is given by
\bea \label{ldl=}
\lambda_p{}^\alpha \text{d} \lambda_\alpha{}^q
=  {\cal A}^{pq}+ \frac{1}{2}
 \Omega^{I} (S^T\gamma^{I}S)_{pq},
\qquad
\eea
where $\lambda_p{}^\alpha$ is the matrix inverse to  $ \lambda_\alpha{}^q$,
\bea\label{l-1=v-1S}
\lambda_p{}^\alpha= v_q^{\alpha} S^{qp} , \qquad  v_q^{\alpha} =S^{qp}\lambda_p{}^\alpha \; ,\qquad
\eea
and
\bea\label{cA=}
\qquad {\cal A}^{pq} =-{\cal A}^{qp}= S^{lp}DS^{lq}= \left(S^{-1}dS - \frac{1}{4} \Omega^{IJ}\, S^T\gamma^{IJ}S\right)_{pq}
\eea
is the composite $SO(16)$ connection. In \eqref{cA=} the symbol  $D$ denotes
(as before)
the $SO(8)$ covariant derivative, $DS= dS - \frac{1}{4} \Omega^{IJ}\, \gamma^{IJ}S$. Due to its presence, the $ {\cal A}^{pq}$ (carrying two $SO(16)$ indices) is invariant under local $SO(8)$ symmetry acting on $S^{qp}$ from the left. (Notice that
$SO(16)$ acts on $S^{qp}$ from the right).

The covariant derivatives of the constrained spinor helicity formalism can be obtained by decomposing the differential in the space of constrained helicity spinors on the basis of
$ {\cal A}^{pq}$ and $\Omega^I$ Cartan forms,
\be
d^{(\lambda)}:=d\lambda_\alpha{}^p \frac \partial {\partial \lambda_\alpha{}^p}= \frac 1 2  {\cal A}^{pq} {\cal D}^{pq}+
 \Omega^I{\cal D}^{I}\; .
\ee
They read
\bea\label{cDpq=}
 {\cal D}^{pq}&=& \lambda_\alpha{}^p \frac \partial {\partial \lambda_\alpha{}^q}-\lambda_\alpha{}^q \frac \partial {\partial \lambda_\alpha{}^p}\; , \qquad \\ \label{cDI=}
 {\cal D}^{I}&=&\frac 1 4  \, \left(S^T\gamma^{I} S\right)_{pq}\left(\lambda_\alpha{}^p \frac \partial {\partial \lambda_\alpha{}^q}+\lambda_\alpha{}^q \frac \partial {\partial \lambda_\alpha{}^p}\right) \; , \qquad
\eea
and, as it is not difficult to check, give zero when applied to the constraints \eqref{u0s=ll} and \eqref{ltsl=}.

\bigskip

Let us stress that in the description of our dynamical system,
D$0$--brane, by the action with Lagrangian form \eqref{eq:LD0=}, the dynamical variables are helicity spinors $\lambda_\alpha{}^q$ constrained by  \eqref{u0s=ll} and \eqref{ltsl=}, while the  Stueckelberg fields $S^{pq}$ is a part of it appearing in these constraints. Explicitly it can be defined as

\be\label{S=vl}
S^{qp}= v_q^\alpha\lambda_\alpha{}^p
\ee
where $v_q^\alpha$ is the (inverse to the) spinor moving frame matrix related to the moving frame vectors $u_\mu^0$ and  $u_\mu^I$, which enter \eqref{u0s=ll} and \eqref{ltsl=}, by constraints
\eqref{u0s=vv} and \eqref{vtsv=}.

Gauging $S^{pq}$  to unit $16 \times 16$ matrix $\delta^{pq}$ reduces the constraints  \eqref{u0s=ll} and \eqref{ltsl=} to the constraints  \eqref{u0s=vv} and \eqref{vtsv=} for spinor moving frame matrix and the $SO(16)$ Cartan form \eqref{cA=} to the spinor representation of the $SO(9)$ Cartan form
$\Omega^{IJ}$. The advantage of using the constrained helicity spinor variables is  that the set of covariant momenta, the counterpart of covariant derivatives, will include the classical counterpart of the covariant derivative
\eqref{cDpq=} representing the $SO(16)$ symmetry generator. This will simplify essentially the formulation of the supertwistor approach and its possible future applications.

\section{Properties of the $SO(16) $ bridge variables $ {\rm w}_q^{{A}}$}
\label{app:rmW}

The relations of the  bridge variables $(\bar{{\rm w}}_{qA}, \, {\rm w}_q^{{A}})$
(obeying \eqref{brwrw=I} and thus providing a complex representation for $SO(16)$ valued matrix)
with SO(9) vector frame \eqref{UinSO9} read

\begin{eqnarray} \label{wgIw=UcU}
  && \bar{{\rm w}}_{qA}(S^T\gamma^{I}S)_{qp}\bar{{\rm w}}_{pB}= U_I {\cal U}_{AB}\; , \qquad
 {\rm w}_{q}^{A}(S^T\gamma^{I}S)_{qp}{\rm w}_{p}^{B}= \bar{U}_I \bar{{\cal U}}^{AB}\; , \qquad \nonumber \\
  && \bar{{\rm w}}_{qA}(S^T\gamma^{I}_{qp}S){{\rm w}}_{p}^{B}= iU_I ^{\check{J}}(\sigma^{\check{J}}\bar{{\cal U}})_{A}{}^{B}\;  , \qquad  \\  \nonumber \\
   \label{UIgI=}
&&  U_I (S^T\gamma^IS)_{qp} = 2\bar{{\rm w}}_{qA} \bar{{\cal U}}{}^{AB} \bar{{\rm w}}_{pB}\; , \qquad
\bar{U}{}_I (S^T\gamma^IS)_{qp} = 2{{\rm w}}_{q}^{A} {{\cal U}}_{AB}{\rm w}_p^{B} \; , \qquad
% \\  \label{UIhJgI=}&&
 \nonumber \\
 && U_I^{\check{J}} (S^T\gamma^IS)_{qp} =
2i {{\rm w}}_{(q}^{A} (\sigma^{\check{J}}\bar{{\cal U}}){}_A{}^{B }\bar{{\rm w}}_{p)B} \; .  \qquad
\end{eqnarray}
Here $\sigma^{\check{I}}_{AB}= -\sigma^{\check{I}}_{BA}$
 and \be\label{tsI=-sI}
\tilde{\sigma}{}^{\check{J}AB}= -(\sigma^{\check{J}}_{AB})^*=\; (\sigma^{\check{J}}_{BA})^\dagger\;
\ee
are obtained from $SO(7)$ Clebsch-Gordan coefficients $\sigma^{\check{I}}_{\tilde{A}\tilde{B}}= -\sigma^{\check{I}}_{\tilde{B}\tilde{A}}$  by
$SU(8)$ transformations in its anti-fundamental and fundamental representation, respectively.  They obey

\be\label{sts=SU}
\sigma^{\check{I}}_{AC}\tilde{\sigma}{}^{\check{J}\; CB} + \sigma^{\check{J}}_{AC}\tilde{\sigma}{}^{\check{I}\; CB}= 2 \delta^{\check{I}\check{J}}\delta_A{}^B
; .
\ee

The constraints for the $Spin(9)$ spinor frame variables  $(\bar{\tilde{w}}_{q\tilde{A}}, \, \tilde{w}_q^{\tilde{A}})$, presented in \cite{Bandos:2017eof}, can be  obtained
from \eqref{wgIw=UcU} and \eqref{UIgI=} by setting $S^{qp}=\delta^{qp}$,
${\cal U}^{AB}\mapsto \delta_{\tilde{A}\tilde{B}}$,
$\bar{{\cal U}}^{AB}\mapsto \delta_{\tilde{A}\tilde{B}}$
and $\tilde{\sigma}{}^{\check{J}AB}\mapsto  -\sigma^{\check{J}}_{\tilde{A}\tilde{B}}$.

The action of the covariant derivatives \eqref{bDchI=} on the bridge variables,
\be\label{bbDIw=}
{\bb D}^{\check{I}}\bar{{\rm w}}_{qA}=-\frac i 2 \, {\rm w}_q^B\sigma^{\check{I}}_{BA} =  \left(\bar{{\bb D}}^{\check{I}}{\rm w}_q^A \right)^*
\; , \qquad  \nonumber \\
{\bb D}^{\check{I}}{\rm w}_q^A = -\frac i 2 \, \bar{{\rm w}}_{qB} \tilde{\sigma}^{\check{I} \,BA}=\left(\bar{{\bb D}}^{\check{I}}\bar{{\rm w}}_{qA}\right)^* \; , \qquad etc.,
\ee
can be reproduced using the $SO(16)$ rotations and $SU(8)$ transformations from their action on $Spin(9)$ variables which can be found in  \cite{Bandos:2017eof}.

\end{widetext}
\providecommand{\href}[2]{#2}\begingroup\raggedright\endgroup


\begin{thebibliography}{10}

\bibitem{Bergshoeff:1996tu}
E.~Bergshoeff and P.~K. Townsend, {\slshape {Super D-branes},}
  \href{http://dx.doi.org/10.1016/S0550-3213(97)00072-2}{{\em Nucl. Phys. B}
  {\bfseries 490} (1997) 145--162},
  \href{http://arxiv.org/abs/hep-th/9611173}{{ arXiv:hep-th/9611173}}.

\bibitem{Witten:1995im}
E.~Witten, {\slshape {Bound states of strings and p-branes},}
  \href{http://dx.doi.org/10.1016/0550-3213(95)00610-9}{{\em Nucl. Phys. B}
  {\bfseries 460} (1996) 335--350},
  \href{http://arxiv.org/abs/hep-th/9510135}{{ arXiv:hep-th/9510135}}.

\bibitem{Bandos:2025pxv}
I.~Bandos, U.~D.~M. Sarraga, and M.~Tsulaia, {\slshape {Quantum state spectrum
  and field theory of D0-brane},}
  \href{http://dx.doi.org/10.1007/JHEP12(2025)003}{{\em JHEP} {\bfseries 12}
  (2025) 003}, \href{http://arxiv.org/abs/2509.11324}{{
  arXiv:2509.11324~[hep-th]}}.

\bibitem{Kallosh:1997nr}
R.~Kallosh, {\slshape {Covariant quantization of D-branes},}
  \href{http://dx.doi.org/10.1103/PhysRevD.56.3515}{{\em Phys. Rev. D}
  {\bfseries 56} (1997) 3515--3522},
  \href{http://arxiv.org/abs/hep-th/9705056}{{ arXiv:hep-th/9705056}}.

\bibitem{Bandos:2000tg}
I.~A. Bandos, {\slshape {Super D0-branes at the endpoints of fundamental
  superstring: An Example of interacting brane system},} in {\em {3rd
  International Workshop on Supersymmetries and Quantum Symmetries}}.
\newblock 1, 2000.
\newblock \href{http://arxiv.org/abs/hep-th/0001150}{{ arXiv:hep-th/0001150}}.

\bibitem{Bandos:2022uoz}
I.~Bandos and U.~D.~M. Sarraga, {\slshape {Complete nonlinear action for a
  supersymmetric multiple D0-brane system},}
  \href{http://dx.doi.org/10.1103/PhysRevD.106.066004}{{\em Phys. Rev. D}
  {\bfseries 106} (2022) 066004}, \href{http://arxiv.org/abs/2204.05973}{{
  arXiv:2204.05973~[hep-th]}}.

\bibitem{Bandos:2022dpx}
I.~Bandos and U.~D.~M. Sarraga, {\slshape {Properties of multiple D0-brane
  systems: 11D origin, equations of motion, and their solutions},}
  \href{http://dx.doi.org/10.1103/PhysRevD.107.086006}{{\em Phys. Rev. D}
  {\bfseries 107} (2023) 086006}, \href{http://arxiv.org/abs/2212.14829}{{
  arXiv:2212.14829~[hep-th]}}.

\bibitem{Ferber:1977qx}
A.~Ferber, {\slshape {Supertwistors and Conformal Supersymmetry},}
  \href{http://dx.doi.org/10.1016/0550-3213(78)90257-2}{{\em Nucl. Phys. B}
  {\bfseries 132} (1978) 55--64}.

\bibitem{Shirafuji:1983zd}
T.~Shirafuji, {\slshape {Lagrangian Mechanics of Massless Particles With
  Spin},} \href{http://dx.doi.org/10.1143/PTP.70.18}{{\em Prog. Theor. Phys.}
  {\bfseries 70} (1983) 18--35}.

\bibitem{Eisenberg:1988nt}
Y.~Eisenberg and S.~Solomon, {\slshape {The Twistor Geometry of the Covariantly
  Quantized Brink-Schwarz Superparticle},}
  \href{http://dx.doi.org/10.1016/0550-3213(88)90337-9}{{\em Nucl. Phys. B}
  {\bfseries 309} (1988) 709--732}.

\bibitem{Bandos:1999qf}
I.~A. Bandos, J.~Lukierski, and D.~P. Sorokin, {\slshape {Superparticle models
  with tensorial central charges},}
  \href{http://dx.doi.org/10.1103/PhysRevD.61.045002}{{\em Phys. Rev. D}
  {\bfseries 61} (2000) 045002}, \href{http://arxiv.org/abs/hep-th/9904109}{{
  arXiv:hep-th/9904109}}.

\bibitem{Bandos:2005mb}
I.~Bandos, X.~Bekaert, J.~A. de~Azcarraga, D.~Sorokin, and M.~Tsulaia,
  {\slshape {Dynamics of higher spin fields and tensorial space},}
  \href{http://dx.doi.org/10.1088/1126-6708/2005/05/031}{{\em JHEP} {\bfseries
  05} (2005) 031}, \href{http://arxiv.org/abs/hep-th/0501113}{{
  arXiv:hep-th/0501113}}.

\bibitem{Brandhuber:2008pf}
A.~Brandhuber, P.~Heslop, and G.~Travaglini, {\slshape {A Note on dual
  superconformal symmetry of the N=4 super Yang-Mills S-matrix},}
  \href{http://dx.doi.org/10.1103/PhysRevD.78.125005}{{\em Phys. Rev. D}
  {\bfseries 78} (2008) 125005}, \href{http://arxiv.org/abs/0807.4097}{{
  arXiv:0807.4097~[hep-th]}}.

\bibitem{Arkani-Hamed:2008gz}
N.~Arkani-Hamed, F.~Cachazo, and J.~Kaplan, {\slshape {What is the Simplest
  Quantum Field Theory?},}
  \href{http://dx.doi.org/10.1007/JHEP09(2010)016}{{\em JHEP} {\bfseries 09}
  (2010) 016}, \href{http://arxiv.org/abs/0808.1446}{{
  arXiv:0808.1446~[hep-th]}}.

\bibitem{Bern:2011qn}
Z.~Bern, J.~J. Carrasco, L.~J. Dixon, H.~Johansson, and R.~Roiban, {\slshape
  {Amplitudes and Ultraviolet Behavior of N = 8 Supergravity},}
  \href{http://dx.doi.org/10.1002/prop.201100037}{{\em Fortsch. Phys.}
  {\bfseries 59} (2011) 561--578}, \href{http://arxiv.org/abs/1103.1848}{{
  arXiv:1103.1848~[hep-th]}}.

\bibitem{Elvang:2015rqa}
H.~Elvang and Y.-t. Huang, {\em {Scattering Amplitudes in Gauge Theory and
  Gravity}}.
\newblock Cambridge University Press, 4, 2015.

\bibitem{Penrose:1967wn}
R.~Penrose, {\slshape {Twistor algebra},}
  \href{http://dx.doi.org/10.1063/1.1705200}{{\em J. Math. Phys.} {\bfseries 8}
  (1967) 345}.

\bibitem{Penrose:1972ia}
R.~Penrose and M.~A.~H. MacCallum, {\slshape {Twistor theory: An Approach to
  the quantization of fields and space-time},}
  \href{http://dx.doi.org/10.1016/0370-1573(73)90008-2}{{\em Phys. Rept.}
  {\bfseries 6} (1972) 241--316}.

\bibitem{Penrose:1986ca}
R.~Penrose and W.~Rindler,
  \href{http://dx.doi.org/10.1017/CBO9780511524486}{{\em {SPINORS AND
  SPACE-TIME. VOL. 2: SPINOR AND TWISTOR METHODS IN SPACE-TIME GEOMETRY}}}.
\newblock Cambridge Monographs on Mathematical Physics. Cambridge University
  Press, 4, 1988.

\bibitem{Fedoruk:2003td}
S.~Fedoruk and V.~G. Zima, {\slshape {Bitwistor formulation of massive spinning
  particle},} J. Kharkiv Univ. 585 (2003) 39-48 
  %\href{http://arxiv.org/abs/hep-th/0308154}
  [arXiv:hep-th/0308154].

\bibitem{Bette:2004ip}
A.~Bette, J.~A. de~Azcarraga, J.~Lukierski, and C.~Miquel-Espanya, {\slshape
  {Massive relativistic particle model with spin and electric charge from two
  twistor dynamics},}
  \href{http://dx.doi.org/10.1016/j.physletb.2004.06.051}{{\em Phys. Lett. B}
  {\bfseries 595} (2004) 491--497},
  \href{http://arxiv.org/abs/hep-th/0405166}{{ arXiv:hep-th/0405166}}.

\bibitem{Fedoruk:2005ks}
S.~Fedoruk, A.~Frydryszak, J.~Lukierski, and C.~Miquel-Espanya, {\slshape
  {Extension of the Shirafuji model for massive particles with spin},}
  \href{http://dx.doi.org/10.1142/S0217751X06031703}{{\em Int. J. Mod. Phys. A}
  {\bfseries 21} (2006) 4137--4160},
  \href{http://arxiv.org/abs/hep-th/0510266}{{ arXiv:hep-th/0510266}}.

\bibitem{deAzcarraga:2005ky}
J.~A. de~Azcarraga, A.~Frydryszak, J.~Lukierski, and C.~Miquel-Espanya,
  {\slshape {Massive relativistic particle model with spin from free
  two-twistor dynamics and its quantization},}
  \href{http://dx.doi.org/10.1103/PhysRevD.73.105011}{{\em Phys. Rev. D}
  {\bfseries 73} (2006) 105011}, \href{http://arxiv.org/abs/hep-th/0510161}{{
  arXiv:hep-th/0510161}}.

\bibitem{Fedoruk:2005bs}
S.~Fedoruk, {\slshape {Twistor transform for spinning particle},}
  \href{http://dx.doi.org/10.1063/1.1923332}{{\em AIP Conf. Proc.} {\bfseries
  767} (2005) 106--126}.

\bibitem{deAzcarraga:2008ik}
J.~A. de~Azcarraga, J.~M. Izquierdo, and J.~Lukierski, {\slshape
  {Supertwistors, massive superparticles and k-symmetry},}
  \href{http://dx.doi.org/10.1088/1126-6708/2009/01/041}{{\em JHEP} {\bfseries
  01} (2009) 041}, \href{http://arxiv.org/abs/0808.2155}{{
  arXiv:0808.2155~[hep-th]}}.

\bibitem{deAzcarraga:2014hda}
J.~A. de~Azcarraga, S.~Fedoruk, J.~M. Izquierdo, and J.~Lukierski, {\slshape
  {Two-twistor particle models and free massive higher spin fields},}
  \href{http://dx.doi.org/10.1007/JHEP04(2015)010}{{\em JHEP} {\bfseries 04}
  (2015) 010}, \href{http://arxiv.org/abs/1409.7169}{{
  arXiv:1409.7169~[hep-th]}}.

\bibitem{Uvarov:2019vmd}
D.~V. Uvarov, {\slshape {Multitwistor mechanics of massless superparticle on
  AdS5 x S5 superbackground},}
  \href{http://dx.doi.org/10.1016/j.nuclphysb.2019.114830}{{\em Nucl. Phys. B}
  {\bfseries 950} (2020) 114830}, \href{http://arxiv.org/abs/1907.13613}{{
  arXiv:1907.13613~[hep-th]}}.

\bibitem{Kim:2021rda}
J.-H. Kim, J.-W. Kim, and S.~Lee, {\slshape {The relativistic spherical top as
  a massive twistor},} \href{http://dx.doi.org/10.1088/1751-8121/ac11be}{{\em
  J. Phys. A} {\bfseries 54} (2021) 335203},
  \href{http://arxiv.org/abs/2102.07063}{{ arXiv:2102.07063~[hep-th]}}.

\bibitem{Kim:2026opo}
J.-H. Kim, {\slshape {The Kerr two-twistor particle},}
  \href{http://arxiv.org/abs/2602.19495}{{ arXiv:2602.19495~[gr-qc]}}.

\bibitem{Kim:2026yqo}
J.-H. Kim, {\slshape {The Kerr-Newman two-twistor particle},}
  \href{http://arxiv.org/abs/2603.07537}{{ arXiv:2603.07537~[gr-qc]}}.

\bibitem{Dittmaier:1998nn}
S.~Dittmaier, {\slshape {Weyl-van der Waerden formalism for helicity amplitudes
  of massive particles},}
  \href{http://dx.doi.org/10.1103/PhysRevD.59.016007}{{\em Phys. Rev. D}
  {\bfseries 59} (1998) 016007}, \href{http://arxiv.org/abs/hep-ph/9805445}{{
  arXiv:hep-ph/9805445}}.

\bibitem{Conde:2016izb}
E.~Conde, E.~Joung, and K.~Mkrtchyan, {\slshape {Spinor-Helicity Three-Point
  Amplitudes from Local Cubic Interactions},}
  \href{http://dx.doi.org/10.1007/JHEP08(2016)040}{{\em JHEP} {\bfseries 08}
  (2016) 040}, \href{http://arxiv.org/abs/1605.07402}{{
  arXiv:1605.07402~[hep-th]}}.

\bibitem{Arkani-Hamed:2017jhn}
N.~Arkani-Hamed, T.-C. Huang, and Y.-t. Huang, {\slshape {Scattering amplitudes
  for all masses and spins},}
  \href{http://dx.doi.org/10.1007/JHEP11(2021)070}{{\em JHEP} {\bfseries 11}
  (2021) 070}, \href{http://arxiv.org/abs/1709.04891}{{
  arXiv:1709.04891~[hep-th]}}.

\bibitem{Bandos:2026wyk}
I.~Bandos and M.~Tsulaia, {\slshape {Spinor moving frame, type II superparticle
  quantization, hidden $SU(8)$ symmetry of linearized 10D supergravity, and
  superamplitudes},} \href{http://arxiv.org/abs/2603.06404}{{
  arXiv:2603.06404~[hep-th]}}.

\bibitem{Bandos:1990ji}
I.~A. Bandos, {\it Superparticle in Lorentz harmonic superspace}. 
%(In  Russian)},} 
{\em Sov. J. Nucl. Phys.} {\bf 51} (1990) 906--914.

\bibitem{Bandos:2006nr}
I.~A. Bandos, J.~A. de~Azcarraga, and D.~P. Sorokin, {\slshape {On D=11
  supertwistors, superparticle quantization and a hidden SO(16) symmetry of
  supergravity},} in {\em {22nd Max Born Symposium on Quantum, Super and
  Twistors: A Conference in Honor of Jerzy Lukierski on His 70th Birthday}}.
\newblock 12, 2006.
\newblock \href{http://arxiv.org/abs/hep-th/0612252}{{ arXiv:hep-th/0612252}}.

\bibitem{Green:1999by}
M.~B. Green, M.~Gutperle, and H.~H. Kwon, {\slshape {Light cone quantum
  mechanics of the eleven-dimensional superparticle},}
  \href{http://dx.doi.org/10.1088/1126-6708/1999/08/012}{{\em JHEP} {\bfseries
  08} (1999) 012}, \href{http://arxiv.org/abs/hep-th/9907155}{{
  arXiv:hep-th/9907155}}.

\bibitem{Bandos:1992np}
I.~A. Bandos and A.~A. Zheltukhin, {\slshape {Green-Schwarz superstrings in
  spinor moving frame formalism},}
  \href{http://dx.doi.org/10.1016/0370-2693(92)91957-B}{{\em Phys. Lett. B}
  {\bfseries 288} (1992) 77--84}.

\bibitem{Bandos:1992ze}
I.~A. Bandos and A.~A. Zheltukhin, {\slshape {Twistor-like approach in the
  Green-Schwarz D=10 superstring theory},} {\em Phys. Part. Nucl.} {\bfseries
  25} (1994) 453--477.

\bibitem{Bandos:1992hu}
I.~A. Bandos and A.~A. Zheltukhin, {\slshape {Generalization of Newman-Penrose
  dyads in connection with the action integral for supermembranes in an
  eleven-dimensional space},} {\em JETP Lett.} {\bfseries 55} (1992) 81--84.

\bibitem{Bandos:1993yc}
I.~A. Bandos and A.~A. Zheltukhin, {\slshape {Eleven-dimensional supermembrane
  in a spinor moving repere formalism},}
  \href{http://dx.doi.org/10.1142/S0217751X93000424}{{\em Int. J. Mod. Phys. A}
  {\bfseries 8} (1993) 1081--1092}.

\bibitem{Bandos:1994eu}
I.~A. Bandos and A.~A. Zheltukhin, {\slshape {N=1 superp-branes in twistor -
  like Lorentz harmonic formulation},}
  \href{http://dx.doi.org/10.1088/0264-9381/12/3/002}{{\em Class. Quant. Grav.}
  {\bfseries 12} (1995) 609--626}, \href{http://arxiv.org/abs/hep-th/9405113}{{
  arXiv:hep-th/9405113}}.

\bibitem{Bandos:2006wb}
I.~Bandos and D.~Sorokin, {\slshape {Aspects of D-brane dynamics in
  supergravity backgrounds with fluxes, kappa-symmetry and equations of motion:
  Part IIB},} \href{http://dx.doi.org/10.1016/j.nuclphysb.2006.10.010}{{\em
  Nucl. Phys. B} {\bfseries 759} (2006) 399--446},
  \href{http://arxiv.org/abs/hep-th/0607163}{{ arXiv:hep-th/0607163}}.

\bibitem{Vasiliev:2001zy}
M.~A. Vasiliev, {\slshape {Conformal higher spin symmetries of 4-d massless
  supermultiplets and osp(L,2M) invariant equations in generalized
  (super)space},} \href{http://dx.doi.org/10.1103/PhysRevD.66.066006}{{\em
  Phys. Rev. D} {\bfseries 66} (2002) 066006},
  \href{http://arxiv.org/abs/hep-th/0106149}{{ arXiv:hep-th/0106149}}.

\bibitem{Plyushchay:2003gv}
M.~Plyushchay, D.~Sorokin, and M.~Tsulaia, {\slshape {Higher spins from
  tensorial charges and OSp(N|2n) symmetry},}
  \href{http://dx.doi.org/10.1088/1126-6708/2003/04/013}{{\em JHEP} {\bfseries
  04} (2003) 013}, \href{http://arxiv.org/abs/hep-th/0301067}{{
  arXiv:hep-th/0301067}}.

\bibitem{Bandos:2004nn}
I.~Bandos, P.~Pasti, D.~Sorokin, and M.~Tonin, {\slshape {Superfield theories
  in tensorial superspaces and the dynamics of higher spin fields},}
  \href{http://dx.doi.org/10.1088/1126-6708/2004/11/023}{{\em JHEP} {\bfseries
  11} (2004) 023}, \href{http://arxiv.org/abs/hep-th/0407180}{{
  arXiv:hep-th/0407180}}.

\bibitem{Sorokin:2017irs}
D.~Sorokin and M.~Tsulaia, {\slshape {Higher Spin Fields in Hyperspace. A
  Review},} \href{http://dx.doi.org/10.3390/universe4010007}{{\em Universe}
  {\bfseries 4} (2018) 7}, \href{http://arxiv.org/abs/1710.08244}{{
  arXiv:1710.08244~[hep-th]}}.

\bibitem{Vasiliev:1999ba}
M.~A. Vasiliev, {\slshape {Higher spin gauge theories: Star product and AdS
  space},} \href{http://arxiv.org/abs/hep-th/9910096}{{ arXiv:hep-th/9910096}}.

\bibitem{Bandos:2001pu}
I.~A. Bandos, J.~A. de~Azcarraga, J.~M. Izquierdo, and J.~Lukierski, {\slshape
  {BPS states in M theory and twistorial constituents},}
  \href{http://dx.doi.org/10.1103/PhysRevLett.86.4451}{{\em Phys. Rev. Lett.}
  {\bfseries 86} (2001) 4451--4454},
  \href{http://arxiv.org/abs/hep-th/0101113}{{ arXiv:hep-th/0101113}}.

\bibitem{Bandos:2002te}
I.~A. Bandos, {\slshape {BPS preons and tensionless superp-brane in generalized
  superspace},} \href{http://dx.doi.org/10.1016/S0370-2693(03)00272-7}{{\em
  Phys. Lett. B} {\bfseries 558} (2003) 197--204},
  \href{http://arxiv.org/abs/hep-th/0208110}{{ arXiv:hep-th/0208110}}.

\bibitem{Bandos:2006xz}
I.~A. Bandos, J.~A. de~Azcarraga, and O.~Varela, {\slshape {On the absence of
  BPS preonic solutions in IIA and IIB supergravities},}
  \href{http://dx.doi.org/10.1088/1126-6708/2006/09/009}{{\em JHEP} {\bfseries
  09} (2006) 009}, \href{http://arxiv.org/abs/hep-th/0607060}{{
  arXiv:hep-th/0607060}}.

\bibitem{Bandos:2008um}
I.~A. Bandos and J.~A. de~Azcarraga, {\slshape {BPS preons and the
  AdS-M-algebra},} \href{http://dx.doi.org/10.1088/1126-6708/2008/04/069}{{\em
  JHEP} {\bfseries 04} (2008) 069}, \href{http://arxiv.org/abs/0802.2890}{{
  arXiv:0802.2890~[hep-th]}}.

\bibitem{Casalbuoni:1976tz}
R.~Casalbuoni, {\slshape {The Classical Mechanics for Bose-Fermi Systems},}
  \href{http://dx.doi.org/10.1007/BF02729860}{{\em Nuovo Cim. A} {\bfseries 33}
  (1976) 389}.

\bibitem{Dirac:1963}
P.~Dirac, {\em {Lectures on Quantum Mechanics}}.
\newblock Snowball Publishing, 2012.

\bibitem{Bandos:2007wm}
I.~A. Bandos, {\slshape {D=11 massless superparticle covariant quantization,
  pure spinor BRST charge and hidden symmetries},}
  \href{http://dx.doi.org/10.1016/j.nuclphysb.2007.12.019}{{\em Nucl. Phys. B}
  {\bfseries 796} (2008) 360--401}, \href{http://arxiv.org/abs/0710.4342}{{
  arXiv:0710.4342~[hep-th]}}.

\bibitem{Berkovits:2002uc}
N.~Berkovits, {\slshape {Towards covariant quantization of the supermembrane},}
  \href{http://dx.doi.org/10.1088/1126-6708/2002/09/051}{{\em JHEP} {\bfseries
  09} (2002) 051}, \href{http://arxiv.org/abs/hep-th/0201151}{{
  arXiv:hep-th/0201151}}.

\bibitem{Green:1987sp}
M.~B. Green, J.~H. Schwarz, and E.~Witten, {\em {Superstring Theory. vol. 1:
  Introduction}}.
\newblock Cambridge Monographs on Mathematical Physics. Cambridge University
  Press, 7, 1988.

\bibitem{Green:1998by}
M.~B. Green and S.~Sethi, {\slshape {Supersymmetry constraints on type IIB
  supergravity},} \href{http://dx.doi.org/10.1103/PhysRevD.59.046006}{{\em
  Phys. Rev. D} {\bfseries 59} (1999) 046006},
  \href{http://arxiv.org/abs/hep-th/9808061}{{ arXiv:hep-th/9808061}}.

\bibitem{Bandos:2017eof}
I.~Bandos, {\slshape {Spinor frame formalism for amplitudes and constrained
  superamplitudes of 10D SYM and 11D supergravity},}
  \href{http://dx.doi.org/10.1007/JHEP11(2018)017}{{\em JHEP} {\bfseries 11}
  (2018) 017}, \href{http://arxiv.org/abs/1711.00914}{{
  arXiv:1711.00914~[hep-th]}}.

\bibitem{Galperin:1992pz}
A.~S. Galperin, P.~S. Howe, and P.~K. Townsend, {\slshape {Twistor transform
  for superfields},} \href{http://dx.doi.org/10.1016/0550-3213(93)90651-5}{{\em
  Nucl. Phys. B} {\bfseries 402} (1993) 531--547}.

\bibitem{Bandos:2017zap}
I.~Bandos, {\slshape {An analytic superfield formalism for tree superamplitudes
  in D=10 and D=11},} \href{http://dx.doi.org/10.1007/JHEP05(2018)103}{{\em
  JHEP} {\bfseries 05} (2018) 103}, \href{http://arxiv.org/abs/1705.09550}{{
  arXiv:1705.09550~[hep-th]}}.

\end{thebibliography}
\end{document}